\documentclass[letterpaper,11pt]{article}
\pdfoutput=1 
\usepackage{jheppub}
\usepackage[utf8]{inputenc} 
\usepackage{graphicx}
\usepackage{amsmath}
\usepackage{amsfonts}
\usepackage{slashed}
\usepackage{amssymb}
\usepackage{hyperref}
\hypersetup{colorlinks=true, linkcolor=blue, citecolor=red, urlcolor=cyan, linktoc=page}
\usepackage{xfrac}
\usepackage{empheq}
\usepackage{caption, subcaption}
\usepackage{parskip}

\newcommand{\p}{\partial}
\newcommand{\oi}{\omega_i}
\newcommand{\oj}{\omega_j}
\newcommand{\ok}{\omega_k}
\newcommand{\ol}{\omega_\ell}
\newcommand{\oone}{\overline{\omega}_1}
\newcommand{\otwo}{\overline{\omega}_2}
\newcommand{\thi}{\theta_i}
\newcommand{\thj}{\theta_j}
\newcommand{\thk}{\theta_k}
\newcommand{\thl}{\theta_\ell}
\newcommand{\mc}{\mathcal}

\newcommand{\ob}{\overline{\omega}}

\newcommand{\obet}{\omega_{\beta}}
\newcommand{\ogam}{\omega_\gamma}

\title{Examining Instabilities Due to Driven Scalars in AdS}

\author{Brad Cownden}
\affiliation{Department of Physics \& Astronomy\\ University of Manitoba,
Winnipeg, Manitoba R3T 2N2, Canada}
\emailAdd{cowndenb@myumanitoba.ca}

\abstract{We extend the study of the non-linear perturbative theory of weakly turbulent energy cascades in AdS$_{d+1}$ to include solutions of driven systems, i.e.~those with time-dependent sources on the AdS boundary. This necessitates the activation of non-normalizable modes in the linear solution for the massive bulk scalar field, which couple to the metric and normalizable scalar modes. We determine analytic expressions for secular terms in the renormalization flow equations mass values $m_{BF}^2 < m^2 \leq 0$, and for various driving functions. Finally, we numerically evaluate these sources for $d=4$ and discuss what role these driven solutions play in the perturbative stability of AdS.}

\keywords{Anti-de Sitter Instability, Gauge/Gravity Duality, Holographic Quantum Quenches}

\begin{document}
\maketitle
\flushbottom
\newpage


\section{Introduction}

Nonlinear instabilities in Anti-de Sitter space have been the subject of examinations on several grounds in addition to  the holographic description of quantum quenches via the AdS/CFT correspondence \cite{ 1708.05600, 1501.00007}, including general stability of maximally-symmetric solutions in general relativity \cite{1104.3702, 1208.5772, 1706.06101}, and the study of the growth of secular terms in time-dependent perturbation theories \cite{hep-th/9506161, 1305.4117}. Numerical studies in AdS show that the eventual collapse of a scalar field into a black hole in the bulk (which is dual to the thermalization of the boundary theory) is generic to any finite sized perturbation \cite{1104.3702, 1106.2339, 1108.4539}, but can be avoided or delayed for certain initial conditions \cite{1803.02830, 1711.00454, 1706.07413, 1508.02709}. The mechanism of collapse in such systems is described as a weakly turbulent energy cascade to short length scales. These dynamics can be captured by a non-linear perturbation theory at first non-trivial order through the introduction of a second, ``slow time'' that describes energy transfer between the fundamental modes. This is known as the Two-Time Formalism (TTF) \cite{1403.6471} and yields renormalization flow equations that allow for the absorption of secular terms into renormalized amplitudes and phases \cite{1308.2132, 1508.05474, 1407.6273, 1412.3249, 1508.04943}. This way, stability against a perturbation of order $\epsilon$ is maintained over time scales of $t \sim \epsilon^{-2}$.

Conventional examinations of perturbative stability using TTF have focused on the reaction of the bulk space to some initial energy perturbation, and have aimed to study the balance between direct and inverse energy cascades \cite{1606.02712, 1612.04758, 1602.05859, 1507.08261, 1607.08094}. Furthermore, numerical examinations of ``pumped'' scalars and their implications for thermalization of the dual theory have also been examined \cite{1410.6201, 1712.07637, 1612.07701, 1502.05726, 1706.02438}. However, extensions of the perturbative description to include time-dependent sources -- corresponding to a driving term on the boundary of the bulk space -- remain unaddressed. 

With this in mind, we examine the effects that a time-dependent source on the conformal boundary have on the non-linear perturbative theory in the bulk. The introduction of a driving term on the boundary means that we must include a second class of fundamental modes with arbitrary frequencies \cite{hep-th/9805171}. Since these solutions will have non-finite inner products over the bulk space, they are known as non-normalizable modes. Non-normalizable modes are identified through the AdS/CFT correspondence with sources coupling to boundary operators \cite{hep-th/9802150, hep-th/9802109}.

Following the TTF prescription, we expand all fields in powers of a small perturbation $\epsilon$. At third order in this perturbation, we find expressions for the resonant terms that are responsible for the weakly turbulent transfer of energy to short length scales which will eventually lead to gravitational collapse. The form of the resonant terms depends on the specific physics of the system, as well as possible symmetries between frequencies. While in general types of resonances are able to be removed by frequency shifts, combinations of frequencies that satisfy a resonance condition will produce secular growth that cannot be absorbed by simple frequency shifts. These terms will be absorbed into the definition of slowly-varying integration constants through a set of renormalization flow equations so that the scalar field remains stable against collapse over perturbative timescales.

This paper is organized as follows: section \S\!~\ref{sec: source terms and BCs} discusses how the presence of a time-dependent boundary condition on the scalar field results in the activation of non-normalizable modes. We briefly discuss the interpretation of non-normalizable modes within the AdS/CFT correspondence and how their behaviour near the boundary is related to energy fluctuations in the CFT. Furthermore, we specify any restrictions on the form of the resonances that we are considering. In section~\S\!~\ref{sec: NNmodes}, we consider a variety of boundary conditions for non-normalizable modes in the bulk. For each choice of boundary condition, we derive analytic expressions for applicable resonances and evaluate these expressions numerically in four spacetime dimensions for different ranges of scalar field masses. Finally, in~\S\!~\ref{sec: discussion} we discuss the implications of non-vanishing resonances on slowly-varying amplitudes and phases, and contrast these results with those for static boundary conditions. Whenever values are calculated, the choice of $d=4$ is implied as to draw the most direct comparison to existing literature such as~\cite{1412.3249, 1407.6273, 1508.04943, 1606.02712}. 

For completeness, we include details of our derivation of the general source term in Appendix~\ref{app: source term derivation} in the boundary time gauge. As an exercise -- and to provide explicit expressions for the resonant contributions when the scalar field has non-zero mass -- Appendix~\ref{sec: norm res} contains the calculation of secular terms in the case of a massive scalar field in AdS$_{d+1}$ with any mass-squared, up to and including the Breitenlohner-Freedman mass \cite{Breitenlohner:1982bm}: $m^2_{BF} \leq m^2$. We present numerical evidence of the natural vanishing of two of the three resonances, and then examine the effects of mass-dependence on the non-vanishing channel. Finally, Appendix~\ref{more 2NN} focuses on additional possible contributions to the source term when the boundary condition is a single non-normalizable mode but with specific values for the frequency.


\section{Normalizable and Non-normalizable Modes}
\label{sec: source terms and BCs}

Let us consider a spherically symmetric, minimally coupled, massive scalar field on an asymptotically AdS$_{d+1}$ spacetime in global coordinates whose metric is given by
\begin{align}
\label{AdS metric}
ds^2 &= \frac{L^2}{\cos(x)} \left( - A(t,x) e^{-2 \delta(t,x)} \, dt^2 + A^{-1}(t, x) \, dx^2 + \sin^2 (x) \, d\Omega^2_{d-1} \right) \, ,
\end{align}
where $L$ is the AdS curvature (hereafter set to $1$), and the radial coordinate $x \in [0, \pi/2)$. The dynamics of the system come from the Einstein and Klein-Gordon equations:
\begin{align}
G_{\mu \nu} + \Lambda g_{\mu \nu} = 8 \pi \left( \nabla_\mu \phi \nabla_\nu \phi - \frac{1}{2} g_{\mu \nu} \left( \nabla^\rho \phi \nabla_\rho \phi + m^2 \phi^2 \right) \right) \quad \text{and} \quad \nabla^2 \phi - m^2 \phi = 0 \, ,
\end{align}
where the cosmological constant $\Lambda$ for AdS given by $\Lambda = -d(d-1)/2$. Perturbing around static AdS, the scalar field is expanded in odd powers of epsilon 
\begin{align}
\phi(t,x) = \epsilon \phi_1(t,x) + \epsilon^3 \phi_3 (t,x) + \ldots
\end{align}
and the metric functions $A$ and $\delta$ in even powers,
\begin{align}
A(t, x) &= 1 + \epsilon^2 A_2 (t,x) + \ldots \\
\delta(t, x) &= \epsilon^2 \delta_2 (t,x)+ \ldots \, .
\end{align}
We choose to work in the boundary gauge, where $\delta(t, \pi/2) = 0$, for reasons that we discuss below.

At linear order, $\phi_1$ satisfies
\begin{align}
\label{phi1 eqn}
\p^2_t \phi_1 + \hat L \phi_1 = 0 \quad \text{and}& \quad \phi_1(t, x \to \pi/2) = \mc F(t) \left( \cos(x) \right)^{\Delta^-}\, \\
\text{where} \;\; \hat{L} &\equiv \frac{1}{\mu} (\mu' \p_x + \mu \p^2_x)  - \frac{m^2}{\cos^2(x)} \, ,
\end{align}
and $\mu \equiv \tan^{d-1}(x)$. The time-dependent boundary condition for $\phi_1$ is the result of the insertion of a \emph{driving term} in the conformal theory that a function only of time. Parameterizing the scalar field by
\begin{align}
	\phi_1 (t, x) = \sum_I c_I (t) E_I (x)
\end{align}
yields an solution to \eqref{phi1 eqn} in the bulk with $c_I(t) = a_I \cos \left( \omega_I t + b_I \right)$ as well as an
eigenvalue equation for the spatial part
\begin{align}
	\label{eigen eqn}
\hat L E_I (x) = \omega_I^2 E_I (x) \, .
\end{align}
Requiring regularity at the origin we find that \cite{hep-th/9805171}
\begin{align}
	\label{general basis}
	E_I (x) =  K_I \left( \cos(x) \right)^{\Delta^+} {_2F_1} \left(\frac{\Delta^+ + \omega_I}{2}, \frac{\Delta^+ - \omega_I}{2}, d/2 ; \sin^2 (x) \right) ,
\end{align}
where the positive (negative) root of $\Delta (\Delta - d) = m^2$ as $\Delta^+$($\Delta^-$). There are two solution for $E_I(x)$ at the boundary, denoted $\Phi^\pm_I (x)$, that combine to give the solution at the origin \eqref{general basis} through $E_I (x) = C_I^+ \Phi^+_I(x) + C^{-}_I \Phi^{-}_I (x)$, where $C^{\pm}_I$ are constants that depend on the frequency $\omega_I$ and scaling dimension $\Delta^{\pm}$. By examining each function's scaling when $x \to \pi/2$,  we see that $\Phi_I^+$ goes as $(\cos x )^{\Delta^+}$ near the boundary and therefore is normalizable; furthermore, $\Phi_I^-$ goes as $( \cos x )^{\Delta^-}$ in this limit and therefore is non-normalizable with respect to the inner product
\begin{align}
	\label{inner prod}
	\langle f(x), g(x) \rangle = \int^{\pi/2}_0 dx \, \mu(x) f^\dagger(x) g(x) \, .
\end{align}

Matching this behaviour with the time-dependent boundary condition in \eqref{phi1 eqn}, we see that the non-normalizable modes are the ones that couple to the driving term \cite{Nastase}. It is important to realize that the general solution for the scalar field in the bulk, \eqref{general basis}, contains both normalizable \emph{and} non-normalizable components.

For special values of the frequencies $\omega_I = \omega_i = 2i + \Delta^+$ with $i \in \mathbb{Z}^+$, the solution is purely normalizable 
and can be written as
\begin{align}
\label{normal basis}
E_I (x)\Big|_{\omega_I = \oi} = e_i(x) &= k_i \left( \cos(x) \right)^{\Delta^+} P_{i}^{(d/2 - 1, \, \Delta^+ - d/2)} \left( \cos (2x) \right) \, ,
\end{align}
with the Jacobi polynomials $P^{(a,b)}_n (x)$ satisfying ${\langle e_i (x), e_j(x) \rangle = \delta_{i j}}$ with respect to the inner product \eqref{inner prod} when
\begin{align}
k_i &= 2 \sqrt{\frac{(i + \Delta^+ /2) \Gamma(i+1) \Gamma(i+\Delta^+)}{\Gamma(i+d/2) \Gamma(i + \Delta^+ - d/2 + 1)}} \, .
\end{align}
For consistency with other frequency values, we choose to write all modes in the general form of \eqref{general basis}. 

Let us now write the first-order part of the scalar field as a sum over both normalizable and non-normalizable modes
\begin{align}
\label{phi1 gen}
\phi_1(t,x) &= \sum_I c_I (t)  E_I (x) \nonumber \\
&= \sum_j a_j (t) \cos \left( \oi t + b_i(t) \right) e_j(x) + \sum_\alpha \bar A_\alpha \cos \left( \omega_\alpha t + \bar{B}_\alpha \right) E_\alpha(x) \, .
\end{align}
As we have seen, the addition of a time-dependent boundary condition on the boundary of AdS activates the non-normalizable modes of the scalar field. The values of $\bar A_\alpha$ and $\bar{B}_\alpha$ will be set by the particular form of the driving term, which we impose. This justifies our choice of working in the boundary gauge; the time $t$ is the proper time measured on the boundary, as well as the time scale of oscillations from the driving term. In the simplest example, the driving term on the boundary is a single, periodic function with (non-integer) frequency $\ob$ and (constant) amplitude $\mc A$
\begin{align}
\label{BC}
\phi_1(t,\pi/2) = \mc A \cos \ob t \, .
\end{align}
In this case, the vanishing of the normalizable modes at the boundary leaves only the non-normalizable part of \eqref{phi1 gen}, which collapses into a single term
\begin{align}
\sum_\alpha \bar A_\alpha \cos (\omega_\alpha t + \bar B_\alpha )  E_{\alpha} (\pi / 2) = \mc A \cos \ob t \:\: \Rightarrow \:\: \bar A_{\ob} \, E_{\ob} (\pi / 2) = \mc A \quad \text{and} \quad \bar B_{\ob} = 0 \, .
\end{align}
In practice, we generalize the boundary condition to a sum over Fourier modes $\ob_\alpha$, which means that additional $\bar A_\alpha$ and $\bar B_\alpha$ terms are non-zero. Note that the allowed non-normalizable frequencies
are completely set by the form of the boundary term being considered. In subsequent sections, we will examine several specific choices of driving frequencies that would produce resonances beyond first order. Note that because the non-normalizable frequencies are not restricted to integer values like the normalizable modes, there are an infinite set of possible boundary configurations that could be explored. Therefore, we will restrict our work to a set of configurations that will be particularly useful in comparing to existing work with driven CFTs \cite{1712.07637, 1502.05726, 1612.07701, 1805.00031}.


\subsection{Interpretation Through the AdS/CFT Dictionary}

The AdS/CFT dictionary relates the leading coefficient of the normalizable modes of the scalar field at the boundary to the expectation value of an operator $\langle \mc O_0 \rangle$ there \cite{hep-th/9808017}. Similarly, the leading coefficient of the non-normalizable modes is related to a source in the boundary CFT. To illustrate this, consider the leading-order behaviour of the scalar field near the conformal boundary:  
\begin{align}
	\phi \sim \alpha (t) \left(\cos (x)\right)^{\Delta^-} + \beta (t) \left( \cos (x) \right)^{\Delta^+} +~\ldots \, ,
\end{align}
where $\alpha(t)$ and $\beta(t)$ are the coefficients of the non-normalizable and normalizable modes, respectively, as determined by their scaling with $x$, and the $(\ldots)$ represent subleading contributions. The Hamiltonian on the boundary receives an extra contribution from the operator $\mc O_0$ that goes like
\begin{align}
	H = H_{CFT} + \alpha(t) \mc O_0 \, .
\end{align}
Therefore, the presence of non-normalizable modes corresponds to a time-dependent portion of the energy of the boundary theory. 

In principle, we can determine the exact form of the time dependence of the energy density in the boundary theory in terms of bulk variables. However, there are additional considerations that arise due to non-normalizable modes. In particular, the boundary CFT must be renormalizable and surface terms from the bulk action may contain contributions from non-normalizable modes that violate renormalizability. The process of holographic renormalization can be approached through the explicit calculation of the required counterterms \cite{hep-th/9804058, hep-th/0112119}, or by rewriting the metric near the boundary in the Fefferman-Graham form \cite{0707.1737, hep-th/0303164, hep-th/0205075, fefferman1985mathematical}. However, this process is complicated by the appearance of different terms when either even or odd numbers of dimensions are considered \cite{hep-th/0205061, 1206.6785}. For these reasons, explicit calculations of one-point functions on the boundary in terms of the bulk variables -- which requires that the relevant counterterms be added -- will be left to future work.

\subsection{The General Source Term}

Without specifying whether frequencies or basis functions have been chosen to be either normalizable or non-normalizable for the time being, we can show that the $\mc O(\epsilon^3)$ part of the scalar field satisfies the equation
\begin{align}
\label{3rd order}
\ddot \phi_3 + \hat L \phi_3 = S = 2 (A_2 - \delta_2) \ddot \phi_1 + (\dot A_2 - \dot \delta_2) \dot\phi_1 + (A_2' -\delta_2' )\phi_1' + m^2 A_2 \phi_1 \sec^2 x \, .
\end{align}
Following the steps outlined in Appendix~\ref{app: source term derivation}, we project \eqref{3rd order} onto the basis of normalizable modes. Since all non-normalizable contributions have been fixed by the $\mc O(\epsilon)$ boundary condition, there will be no contribution from non-normalizable modes at this order. Using the expansion from \eqref{phi1 gen}, we see that ${\langle \phi_3, e_\ell \rangle = c^{(3)}_\ell}$ and so
\begin{align}
\label{S gen}
\ddot c^{(3)}_\ell  + \ol^2 c^{(3)}_\ell = S_\ell \, .
\end{align}
As discussed in \cite{1407.6273}, the solution to \eqref{S gen} will contain terms proportional to 
\begin{align}
\label{cos of thetas}
\cos ( \theta_I \pm \theta_J \pm \theta_K) = \cos \left( (\omega_I \pm \omega_J \pm \omega_K) t + b_I \pm b_J \pm b_K \right) \, .
\end{align}
For choices of $\{\theta_I,\theta_J,\theta_K\}$ such that ${\pm \omega_\ell \neq \pm \omega_I \pm \omega_J \pm \omega_K}$, $S_\ell$ will always be made up of terms of the form in \eqref{cos of thetas}. However, if the frequencies are \emph{resonant}, i.e.~if ${\pm \omega_\ell = \pm \omega_I \pm \omega_J \pm \omega_K}$, then the solution to \eqref{S gen} will contain terms of the form
\begin{align}
\label{secular}
t \sin ( \theta_I \pm \theta_J \pm \theta_K ) \, .
\end{align}
These terms, known as \emph{secular terms}, grow linearly with time and will eventually invalidate the perturbative expansion. While secular terms from non-resonant spectra are able to be absorbed into a shift of $\omega_\ell$, secular terms from fully resonant spectra are not able to be absorbed in this way. Instead, a resummation of secular terms must be applied. Since non-secular terms and secular terms from non-resonant spectra can be controlled for without the use of a resummation scheme, we henceforth let $S_\ell$ denote only secular terms from fully resonant spectra since it is these terms that will determine the evolutions of the first-order amplitudes and phases~\cite{hep-th/9506161}. 

The resummation of secular terms can be achieved through either the Two-Time Formalism picture~\cite{1403.6471} or the renormalization group resummation picture~\cite{1412.3249}. The effect of either method is that we can rewrite the amplitudes and phases in terms of renormalized integration constants that exactly cancel the secular terms at each instant. Doing so yields the renormalization flow equations for the renormalized constants~\cite{1407.6273}
\begin{align}
\label{RN flow 1}
\frac{2 \ol}{\epsilon^2} \frac{d a_\ell}{d t} &= - \sum_{I,J,K} S_\ell \sin \left( b_\ell \pm b_I \pm b_J \pm b_K \right) \\
\label{RN flow 2}
\frac{2 \ol a_\ell}{\epsilon^2} \frac{d b_\ell}{d t} &= - \sum_{I,J,K} S_\ell \cos \left( b_\ell \pm b_I \pm b_J \pm b_K \right) \, ,
\end{align}
where the sums over ${\{I,J,K\}}$ now only include combinations that give resonant spectra. Note that the amplitudes and phases evolve with respect to the ``slow time'' $\tau = \epsilon^2 t$. In practice, once these flow equations can be written down, the perturbative evolution of the system is determined up to a timescale of $t \sim \epsilon^{-2}$.

Returning to the result of projecting \eqref{3rd order} onto the basis of normalizable modes, we find that the source term for the $\ell^{\rm th}$ mode is

\begin{align}
\label{general source}
S_\ell &=\frac{1}{4} \sum_{\substack{I,J,K \\ K \neq \ell}}^\infty \frac{a_I a_J a_K \omega_K}{\ol^2 - \omega_K^2} \bigg[ Z^-_{IJK\ell} (\omega_I + \omega_J - 2\omega_K) \cos (\theta_I + \theta_J - \theta_K) \nonumber \\
&\qquad \frac{}{} - Z^-_{IJK\ell} (\omega_I + \omega_J + 2\omega_K) \cos (\theta_I + \theta_J + \theta_K) + Z^+_{IJK\ell} (\omega_I - \omega_J + 2\omega_K)  \cos(\theta_I - \theta_J + \theta_K) \nonumber \\
& \qquad \frac{}{} - Z^+_{IJK\ell} (\omega_I - \omega_J - 2\omega_K) \cos (\theta_I - \theta_J - \theta_K) \bigg] \nonumber \\
& + \frac{1}{2}\sum_{\substack{I,J,K \\ I \neq J}}^\infty a_I a_J a_K \omega_J \left( H_{IJK\ell} + m^2 V_{JKI\ell} - 2\omega_K^2 X_{IJK\ell} \right) \bigg[ \frac{1}{\omega_I - \omega_J} \big( \cos (\theta_I - \theta_J - \theta_K)  \nonumber \\
& \qquad + \cos(\theta_I - \theta_J + \theta_K) \big) - \frac{1}{\omega_I + \omega_J} \big( \cos (\theta_I + \theta_J - \theta_K)  + \cos ( \theta_I + \theta_J + \theta_K) \big) \bigg] \nonumber \\
& - \frac{1}{4} \sum_{I,J,K}^\infty a_I a_J a_K \bigg[ \left( 2\omega_J \omega_K X_{IJK\ell} + m^2 V_{IJK\ell} \right)\cos(\theta_I + \theta_J - \theta_K) \nonumber \\
& \qquad \frac{}{} -  \left( 2\omega_J\omega_K X_{IJK\ell} - m^2 V_{IJK\ell} \right) \cos(\theta_I - \theta_J - \theta_K) + \left(2\omega_J \omega_K X_{IJK\ell} + m^2 V_{IJK\ell} \right) \cos (\theta_I - \theta_J + \theta_K) \nonumber \\
& \qquad \frac{}{} - \left( 2\omega_J\omega_K X_{IJK\ell} - m^2 V_{IJK\ell} \right) \cos(\theta_I + \theta_J + \theta_K) \bigg] \nonumber \\
& + \frac{1}{4} \sum_{I,J}^\infty a_I a_J a_\ell \ol \Big[ \tilde Z^-_{IJ\ell} (\omega_I + \omega_J - 2\ol) \cos (\theta_I + \theta_J - \thl) - \tilde Z^-_{IJ\ell} (\omega_I + \omega_J + 2\ol) \cos(\theta_I + \theta_J +  \thl) \nonumber \\
& \qquad + \tilde Z^+_{IJ\ell} (\omega_I - \omega_J + 2\ol) \cos(\theta_I - \theta_J + \thl)  - \tilde Z^+_{IJ\ell} (\omega_I - \omega_J - 2\ol) \cos( \theta_I - \theta_J - \thl)  \Big] \nonumber \\
& - \frac{1}{4} \sum_{I,J}^\infty a_I^2 a_J \bigg[ H_{IIJ\ell} + m^2 V_{JII\ell} - 2\omega_J^2 X_{IIJ\ell} \bigg] \big( \cos (2\theta_I - \theta_J) + \cos (2\theta_I + \theta_J) \big) \nonumber \\
& - \frac{1}{2} \sum_{I,J}^\infty a_I^2 a_J \bigg[ H_{IIJ\ell} + m^2 V_{JII\ell} - 2\omega_J^2 X_{IIJ\ell} + 4\omega_I^2 \omega_J^2 P_{J\ell I} + 2\omega_I^2 (M_{J\ell I} + m^2 Q_{J\ell I}) \bigg] \cos \theta_J .
\end{align} 
Note that sums and restrictions on indices must be interpreted as sums and restrictions on \emph{frequencies} when any of the modes is non-normalizable, since $\omega_\alpha \neq 2 \alpha + \Delta^+$ in general.

\subsection{The Resonance Condition}

To determine the form of the secular terms in $S_\ell$, we must consider all combinations of the frequencies ${\{ \omega_I, \omega_J, \omega_K\}}$ that satisfy the resonance condition
\begin{align}
\label{gen res}
\omega_I \pm \omega_J \pm \omega_K = \pm \ol \, .
\end{align}
The allowed values of the non-normalizable frequencies are set by the form of driving term in $\mc F(t)$, i.e. which non-normalizable modes have been excited. For instance, if the driving term is of the form in \eqref{BC}, then $\omega_I = \{\omega_i, \ob\}$. Plugging this restriction into \eqref{gen res}, we could potentially have resonances from
\begin{align}
	\label{one NN}
	\omega_i \pm \omega_j \pm \ob &= \pm \ol \\
	\label{two NN}
	\omega_i \pm \ob \pm \ob &= \pm \ol \\
	\label{three NN}
	\ob \pm \ob \pm \ob &= \pm \ol \, .
\end{align}
Since the frequencies of the normalizable modes are always $\oi = 2i + \Delta^+$, and since the non-normalizable modes have generically non-integer frequencies, \eqref{one NN} and \eqref{three NN} require specific values of $\ob$ to be satisfied. There is only a single non-trivial resonance that can occur without tuning the value of $\ob$, which is $\oi + \ob - \ob = \ol$. As an additional example, consider the choice of ${ \omega_I = \omega_J = \omega_K =\ob = 2n - \Delta^-}$ with $n \in \mathbb{Z}^+$ in \eqref{gen res}. One resonance that arises from the condition \eqref{three NN} is
\begin{align}
\omega_I \pm \omega_J \pm \omega_K = \pm \ol \; \to \; \ob - \ob + \ob = \ol
\end{align}
which -- when working in an even number of dimensions with $d \geq 4$ -- is satisfied for $n = \ell - d/2$. Again, this requires the frequency of the boundary condition to be tuned to a special value. 

In the interest of examining the most generic choices for the driving frequency, we do not consider cases that rely on specially tuned values. In particular, we focus on resonances produced when \emph{only two} of the frequencies in \eqref{gen res} are non-normalizable. Therefore, an important caveat to this work is that it does not present an exhaustive list of possible resonances, and that specific choices for the number of dimensions, mass, and driving frequency could result in cases not addressed here. In fact, tuning the frequency of the boundary condition may lead to some very interesting behaviours which may deserve closer inspection in their own right. In the event that additional resonances are possible, the same procedures used to derive the results in \S\!~\ref{sec: NNmodes} can be applied to more specific scenarios if need be.

The contributions from considering only normalizable modes, when $\{ I, J, K \} = \{i, j ,k\}$, have been considered already in detail for massless scalars in the interior \cite{1407.6273} and boundary \cite{1412.3249} time gauges, as well as massive scalars in the interior time gauge \cite{1810.04753}. We include a detailed derivation of the resonant contributions for a massive scalar when the boundary term is zero -- and therefore only normalizable modes are present -- as an exercise in Appendix~\ref{sec: norm res}. Instead, we will concern ourselves mainly with what new terms arise from the activation of non-normalizable modes while keeping in mind that the the total $\mc O(\epsilon^3)$ source term is always given by the sum of both types of contributions \cite{1810.04753}.

Finally, the definitions of the functions $Z$, $H$, $X$, etc.~in~\eqref{general source} differ slightly from those presented in \cite{1407.6273, 1412.3249} in part because of the gauge choice and in part because of a desire to separate out mass-dependent terms; however, the expressions are made equivalent through applications of integration by parts and setting $m^2 = 0$. To avoid confusion, 
the definitions of $Z$, $H$, $X$, etc. are given explicitly in 
Appendix~\ref{app: source term derivation}.


\section{Resonances From Non-normalizable Modes}
\label{sec: NNmodes}

Now let us consider the excitation of non-normalizable modes by a driving term on the boundary of AdS. Having set $\ol$ to be a normalizable mode, we may ask what restrictions exist on our choices for the other frequencies in \eqref{general source}. As discussed at the end of the previous section, we will focus our attention on resonances that occur when \emph{only two} of $\{ \omega_I, \omega_J, \omega_K \}$ are non-normalizable. Furthermore, we limit the values of the non-normalizable frequencies as little as possible, choosing not to consider frequencies that have been too finely tuned. In such cases there may be many resonance channels that need to be considered, which will add together to constitute the total source term $S_\ell$.

Before proceeding further, it is important to consider what effects the introduction of non-normalizable modes might have on the calculation of the source term $S_\ell$ in \eqref{3rd order}. In particular, since non-normalizable solutions do not have well-defined norms, we do not know \emph{a priori} that the inner products that result from projecting the terms in $S_\ell$ onto the basis of normalizable eigenfunctions are still finite. To investigate this, consider the generic expression for the second-order metric function
\begin{align}
A_2 &= - \nu \int^x_0 dy \, \mu \left( (\dot \phi_1)^2 + (\phi'_1)^2 + m^2 \phi_1^2 \sec^2 x \right) \, ,
\end{align}
in the limit of $x \to \pi/2$, and let the scalar field $\phi_1$ be given by a generic superposition of normalizable and non-normalizable eigenfunctions, as in \eqref{phi1 gen}. Expanding in terms of the small parameter $\tilde x$, and ignoring time-dependent contributions, we find that
\begin{align}
A_2 (\tilde x \equiv \pi /2 - x) \sim \tilde{x}^{-\xi} \left( \frac{2 \tilde{x}^{2+d}}{2 - \xi} - \frac{\tilde{x}^d (1 + \left(\Delta^{-}\right)^2)}{\xi} \right) + \ldots \, ,
\end{align}
where we have defined $\xi = \sqrt{d^2 + 4m^2}$. In the massless case, $\xi = d$ and all powers of $\tilde{x}$ are non-negative; thus, $A_2$ is finite as $\tilde x \to 0$. For tachyonic masses, $m^2_{BF} < m^2 < 0$ so that $0 < \xi < d$ and the limit is again finite. However, when $m^2 > 0$, part of the expression diverges resulting in a non-zero contribution at the conformal boundary. In order for the boundary to remain asymptotically AdS, counter-terms in the bulk action would be required to cancel such divergences -- a case we will not address presently. Furthermore, for masses that saturate the Breitenlohner-Freedman bound, the limit would have to be re-evaluated. We will therefore restrict our discussion to $m^2_{BF} < m^2 \leq 0$ to avoid these issues. A similar check on the near-boundary behaviour of $\delta_2$ shows that the gauge condition ${\delta_2 (t, \pi/2) = 0}$ remains unchanged by the addition of non-normalizable modes given the same restrictions on the mass of the scalar field. With these restrictions in mind, let us now examine the resonances produced by the activation of non-normalizable modes.


\subsection{A Single Non-normalizable Mode}
\label{ssec: equalNN}

As a first case, let us assume that the driving term $\mc F(t)$ is comprised of a single, non-integer frequency component, i.e.~
\begin{align}
\mc F(t) = \bar A_{\ob}\cos \ob t \, ,
\end{align}
where the amplitude of the non-normalizable mode $\bar A_{\ob}$ is constant and fixed by the boundary condition. Resonances from integer values of $\ob$ are addressed separately in Appendix~\ref{more 2NN} as they constitute a special case. Recall that we are considering configurations that satisfy the resonance condition \eqref{gen res} when only one of $\{\omega_I, \omega_I, \omega_K \}$ is normalizable. Thus, accounting for (separate) relabelling among normalizable and non-normalizable indices, the resonance condition is
\begin{align}
\label{gen nn}
\oi - \ob + \ob = \ol \, .
\end{align}
When this resonance condition is met, the remaining normalizable mode will have a frequency equal to $\ol$, collapsing all sums over frequencies so that
\begin{align}
\label{2genNN}
S_\ell = \overline{T}_{\ell} \, a_\ell \bar A_{\ob}^2 \cos (\thl) \, .
\end{align}
 Collecting the appropriate terms in \eqref{general source} and evaluating each possible resonance, we find that
\begin{align}
\label{S:2NN}
\overline{T}_{\ell} &=  \bigg[ \, \frac{1}{2} Z^-_{\ell\ob\ob\ell} \left( \frac{\ob}{\ol + \ob} \right) + \frac{1}{2} Z^+_{\ell\ob\ob\ell} \left( \frac{\ob}{\ol - \ob} \right)  + H_{\ell \ob \ob \ell} \left( \frac{\ob^2}{\ol^2 - \ob^2} \right)  - H_{\ob\ell\ob\ell} \left(\frac{\ol^2}{\ol^2 - \ob^2} \right) \nonumber \\
& - m^2 V_{\ell \ob\ob\ell}  \left(\frac{\ol^2}{\ol^2 - \ob^2} \right) + m^2 V_{\ob\ob\ell\ell} \left( \frac{\ob^2}{\ol^2 - \ob^2} \right) + 2 X_{\ob\ob\ell\ell} \left( \frac{\ob^2 \ol^2}{\ol^2 - \ob^2} \right) - 2 X_{\ell\ell\ob\ob} \left( \frac{\ob^4}{\ol^2 - \ob^2} \right) \bigg]_{\ob \neq \ol} \nonumber \\
&  + \ol^2 X_{\ob\ob\ell\ell}  - \ob^2 X_{\ell\ell\ob\ob} - \frac{3}{2} m^2 V_{\ell\ell\ob\ob} - \frac{1}{2} m^2 V_{\ob\ob\ell\ell}  - \frac{1}{2} H_{\ob\ob\ell\ell} + \ol^2 \tilde{Z}^+_{\ob\ob\ell} - 2 \ob^2 \ol^2 P_{\ell\ell\ob} \nonumber \\
& - \ob^2 \left( \ol^2 P_{\ell\ell \ob} - B_{\ell\ell\ob} \right) \, .
\end{align}

Although the case of $\ob = \ol$ is not permitted physically (since the non-normalizable mode would then have an integer frequency, making it a normalizable mode that does not scale correctly near the boundary), the terms in the square brackets have restrictions on the allowed choices of $\ob$ that come from other sources. For instance, terms that are proportional to $Z^{\pm}$ inherit this restriction from the first sum in \eqref{general source}. The other terms above exactly cancel one another as $\ob \to \ol$, and therefore do not contribute to $\overline{T}_\ell$. For this reason, these terms are grouped with those that have natural restrictions on the indices. 

With the resonant contributions determined, the renormalization flow equations for two equal, constant, non-normalizable frequencies follow from \eqref{RN flow 1}~-~\eqref{RN flow 2} and are
\begin{align}
\label{equal NN flow}
\frac{2 \ol}{\epsilon^2} \frac{d a_\ell}{dt} =  0 \quad \text{and} \quad \frac{2 \ol a_\ell}{\epsilon^2} \frac{d b_\ell}{d t} = - \overline{T}_\ell a_\ell \bar A^2_{\ob} \, .
\end{align}
It is important to reiterate that resonances from normalizable modes are still present whenever non-normalizable modes have been activated, both in this case and all other cases. Therefore \eqref{equal NN flow} must be added to the flow equations for $a_\ell$ and $b_\ell$ from the case of all-normalizable modes. However, it is worth noting that -- in contrast to resonances from all-normalizable modes -- the contribution from non-normalizable modes appears only in the phase term. Indeed, \eqref{equal NN flow} tells us that $b_\ell$ is a linear function of $\tau$ with a slope that is determined by the $\mc O(\epsilon^3)$ physics encapsulated by $\overline{T}_\ell$.

\begin{figure}
\centering
\includegraphics[width=\textwidth]{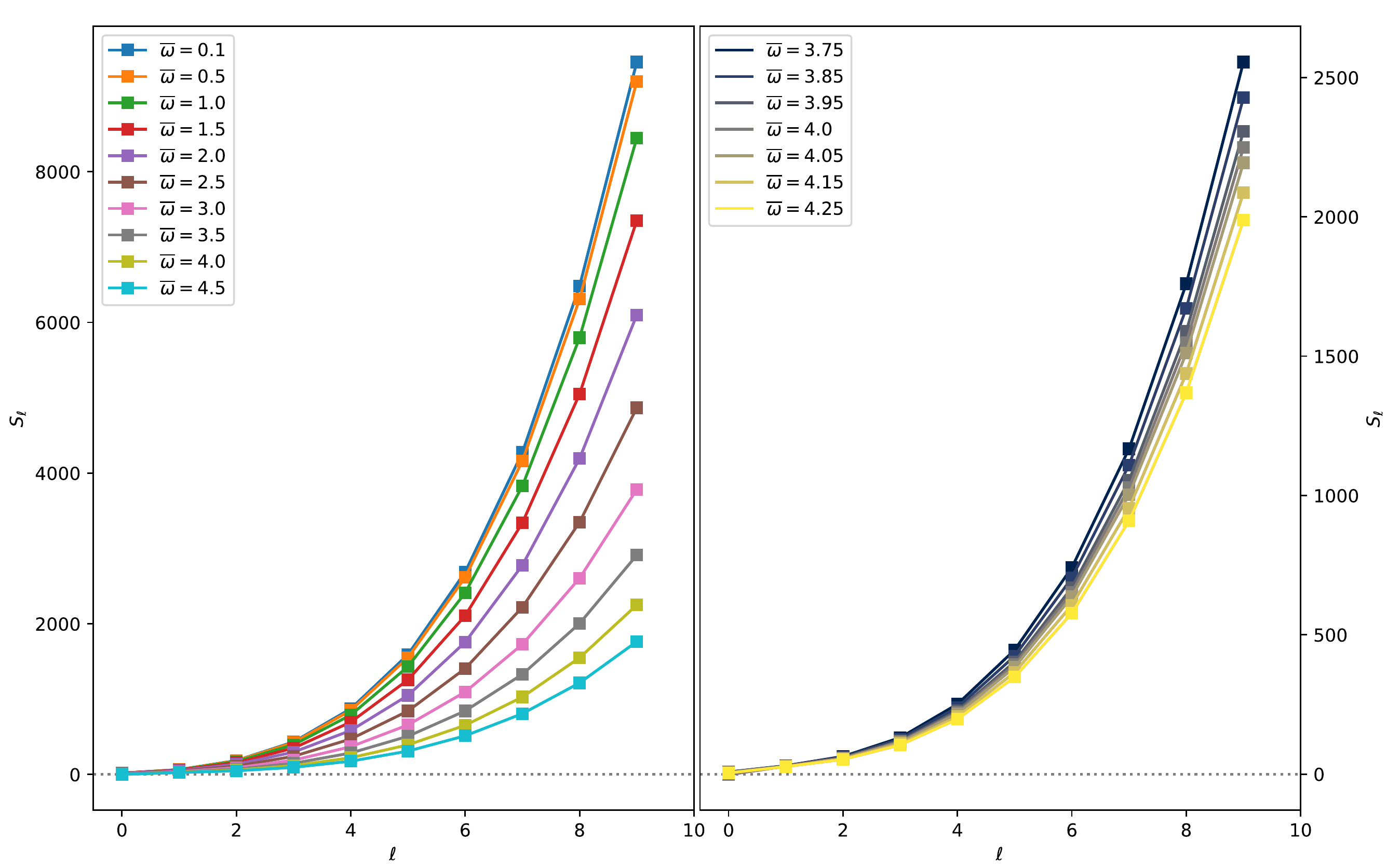}
\caption{{\it Left:} Evaluating \eqref{S:2NN} when $m^2 = 0$ for various choices of $\ob$. {\it Right}: The behaviour of $S_\ell$ for $\ob$ values near $\omega_0$.}
\label{fig:equal_frequency_m0}
\end{figure}

\begin{figure}[h]
\centering
\includegraphics[width=\textwidth]{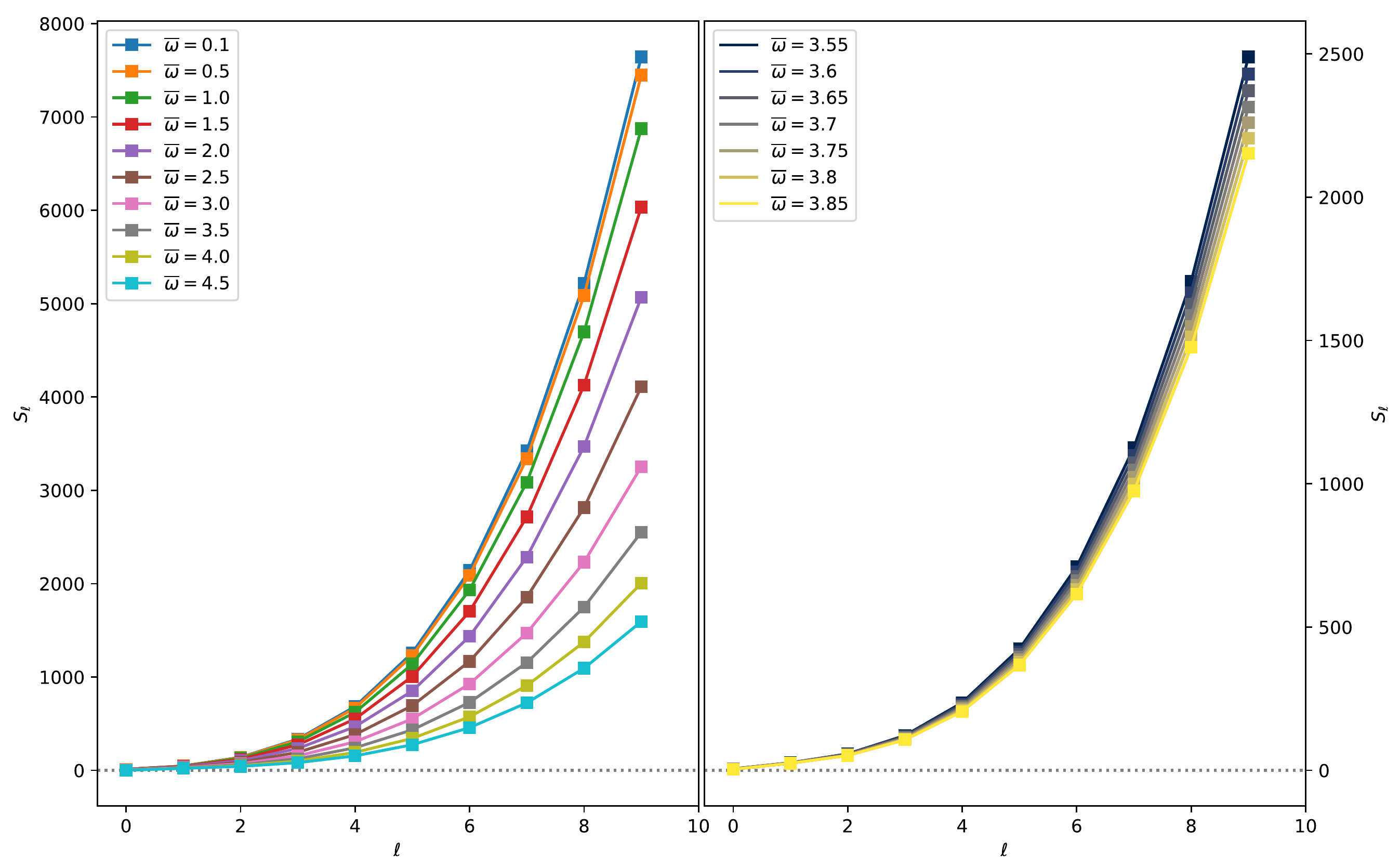}
\caption{{\it Left:} Evaluating $\overline{T}_{\ell}$ for a tachyon with $m^2 = -1.0$. {\it Right:} The behaviour of $S_\ell$ near $\omega_0 = \Delta^+ \approx  3.7$.}
\label{fig:equal_frequency_m-1_0}
\end{figure}

In figures \ref{fig:equal_frequency_m0} and \ref{fig:equal_frequency_m-1_0}, we evaluate \eqref{S:2NN} for $\ell < 10$ over a variety of $\ob$ values first for a massless scalar, then for a tachyonic scalar. For both values of mass-squared, $\overline{T}_\ell$ demonstrates power law-type behaviour as a function of $\ell$ with a leading coefficient that is proportional to the non-normalizable frequency $\ob$. We also see that the limit of \eqref{S:2NN} as $\ob \to \omega_0$ is well-defined in both cases.


\subsection{Special Values of Non-normalizable Frequencies}

Let us now consider special values of non-normalizable frequencies that will lead to a greater number of resonance channels. While general non-normalizable frequencies do not require any such restrictions, we will find it informative to examine these special cases as they possess more symmetry in index/frequency values than the case of equal non-normalizable frequencies, but less than all-normalizable modes. 


\subsubsection{Add to an integer}
\label{ssec: add to integer}

First, we choose the driving term to be given by the sum of two terms whose frequencies $\oone$ and $\otwo$ add to give an integer
\begin{align}
	\mc F(t) = \bar A_1 \cos \oone t + \bar A_2 \cos \otwo t \, ,
\end{align}
where $\oone+ \otwo = 2n$ and $n = 1, 2, 3, \ldots$ (note that the $n = 0$ case means that both $\oone$ and $\otwo$ would need to be zero by the positive-frequency requirement and so would not contribute). Once again, the amplitudes $\bar A_1$ and $\bar A_2$ are constant and fixed by the form of $\mc F(t)$. Furthermore, since neither frequency is restricted to integer values, the difference $|\oone - \otwo|$ will, in general, not be an integer. In \S\!~\ref{ssec: intpluschi}, we examine the case when the difference of non-normalizable frequencies is an integer.

When we consider possible resonance channels, we find two resonances are present for any mass value within $m^2_{BF} < m^2 \leq 0$,
\begin{align}
\label{all pluses}
(++): \; \omega_i + 2n &= \omega_\ell \quad \forall \; \ell \geq n \\
(+-): \, \omega_i - 2n &=\omega_\ell \quad \forall \; n \, ,
\end{align}
while one channel is restricted to massless scalars\footnote{Note that specific values of $m^2 \neq 0$ are capable of producing this kind of resonance channel. In such a case, the condition in \eqref{minus plus} would be $n \geq \ell + \Delta^+$ with $m$, $d$ such that $\Delta^+ \in \; \mathbb{Z}^+$.}
\begin{align}
\label{minus plus}
(-+): \, -\omega_i + 2n = \omega_\ell \quad \forall \; n \geq \ell + d \, .
\end{align}
Furthermore, for all allowed values of $n$ and $m^2$, there are extra contributions due to resonances of the form of \eqref{gen nn}, i.e.
\begin{align}
\label{atoi plus/minus}
	\omega_i + \oone - \oone = \oi + \otwo - \otwo = \ol \quad \forall \; n \, .
\end{align}

Adding the channels together, the total source term is
\begin{align}
\label{add to integer}
S_\ell &=  \!\!\!\! \sum_{\oone + \otwo = 2n}\bigg[ \Theta\left( n - \ell - d \right) \overline{R}^{(-+)}_{(n - \ell - d) \ell} \ \bar A_1 \bar A_2 \, a_{(n - \ell - d)} \cos \left( \theta_{(n - \ell - d)} - \theta_1 - \theta_2 \right) \bigg]_{m^2 = 0}  \nonumber \\ 
&  \quad + \!\!\!\! \sum_{\oone + \otwo = 2n} \!\!\!\! \Theta \left( \ell - n \right)  \overline{R}^{(++)}_{(\ell - n)\ell} \, \bar A_1 \bar A_2 \, a_{(\ell - n)} \cos \left( \theta_{(\ell - n)} + \theta_1 + \theta_2 \right) \nonumber \\
& \quad  + \!\!\!\! \sum_{\oone + \otwo = 2n} \!\!\!\! \overline{R}^{(+-)}_{(\ell + n) \ell} \, \bar A_1 \bar A_2 \, a_{(\ell + n)} \cos\left( \theta_{(\ell + n)} - \theta_1 - \theta_2 \right) \nonumber \\
& \quad + \!\!\!\! \sum_{\oone + \otwo = 2n} \!\!\!\! \overline{U}^{(1)}_{(\ell)\ell} \, \bar A_1^2 \, a_\ell \cos\left( \theta_{(\ell)} + \theta_1 - \theta_1 \right) + \!\!\!\! \sum_{\oone + \otwo = 2n} \!\!\!\! \overline{U}^{(2)}_{(\ell)\ell} \, \bar A_2^2 \, a_\ell \cos\left( \theta_{(\ell)} + \theta_2 - \theta_2 \right) \nonumber \\
& \quad + \!\!\!\! \sum_{\oone + \otwo = 2n} \!\!\!\! \overline{T}^{(1)}_{\ell} \bar A_1^2 \, a_\ell \cos \left( \theta_\ell \right) + \!\!\!\! \sum_{\oone + \otwo = 2n} \!\!\!\! \overline{T}^{(2)}_{\ell} \bar A_2^2 \, a_\ell \cos \left( \theta_\ell \right)
\end{align}
where the Heaviside step function $\Theta(x)$ enforces the restrictions on the indices in \eqref{all pluses} and \eqref{minus plus} and $\theta_1 = \oone t + \bar B_1$, etc. 

In the following expressions, the sum over all $\oone$, $\otwo$ such that $\oone + \otwo = 2n$ is implied, and only the restrictions on individual frequencies are included. Examining each channel in \eqref{add to integer} individually, we find

\begin{align}
\label{R1}
\overline{R}^{(++)}_{i \ell} &= - \frac{1}{4} \sum_{\otwo \neq \ol} \frac{\otwo}{\ol - \otwo} Z^{-}_{i12\ell} - \frac{1}{4} \sum_{\oone \neq \ol} \frac{\oone}{\ol - \oone} Z^{-}_{i21\ell} - \frac{1}{8n} \left( \ol - 2n \right) Z^-_{12i\ell} \nonumber \\
& - \frac{1}{4} \sum_{\oi \neq \oone} \frac{1}{\ol - \otwo} \Big[ \oone \left( H_{i12\ell} + m^2 V_{12i\ell} - 2 \otwo^2 X_{i12\ell} \right) + (\ol - 2n) \left( H_{1i2\ell} + m^2 V_{i21\ell} - 2\otwo^2 X_{1i2\ell} \right)\Big] \nonumber \\
& - \frac{1}{4} \sum_{\oi \neq \otwo} \frac{1}{\ol - \oone} \Big[ \otwo \left( H_{i21\ell} + m^2 V_{21i\ell} - 2\oone^2 X_{i21\ell} \right) + (\ol - 2n) \left( H_{2i1\ell} + m^2 V_{i12\ell} - 2\oone^2 X_{2i1\ell} \right) \Big] \nonumber \\
& - \frac{1}{8n} \sum_{\oone \neq \otwo} \Big[ \oone H_{21i\ell} + \otwo H_{12i\ell} + m^2 \left( \oone V_{1i2\ell} + \otwo V_{2i1\ell} \right) - \left( \ol - 2n \right)^2 \left(\oone X_{21i\ell} + \otwo X_{12i\ell} \right) \Big] \nonumber \\
& + \frac{1}{2} \Big[ \oone\otwo X_{i12\ell} + \left( \ol - 2n \right)\left( \oone X_{21i\ell} + \otwo X_{12i\ell} \right) - \frac{m^2}{2} \left( V_{i12\ell} + V_{i21\ell} + V_{12i\ell} \right) \Big] \, .
\end{align}

The notation $X_{i12\ell}$ corresponds to evaluating $X_{ijk\ell}$ with $\omega_j = \oone$ and $\omega_k = \otwo$. Next, we find that
\begin{align}
\label{R2}
\overline{R}_{i \ell}^{(+-)} &= - \frac{1}{4} \Big[ \frac{(\ol + 2n)}{2n} Z^-_{12i\ell} + 2 (\ol + 2n) \left( \oone X_{21i\ell} + \otwo X_{12i\ell} \right) \nonumber \\
& -\frac{\oone}{(\ol + \otwo)} \left( H_{i12\ell} + m^2 V_{12i\ell} - 2 \otwo^2 X_{i12\ell} \right) + \frac{(\ol + 2n)}{(\ol + \otwo)} \left( H_{1i2\ell} + m^2 V_{i21\ell} - 2\otwo^2 X_{1i2\ell} \right)  \nonumber \\
&- \frac{\otwo}{(\ol + \oone)} \left( H_{i21\ell} + m^2 V_{21i\ell} - 2\oone^2 X_{i21\ell} \right) + \frac{(\ol + 2n)}{(\ol + \oone)} \left(H_{2i1\ell} + m^2 V_{i12\ell} - 2\oone^2 X_{2i1\ell} \right)  \nonumber \\
&  - 2 \oone\otwo X_{i12\ell} + m^2 \left( V_{12i\ell} + V_{i12\ell} + V_{i21\ell} \right) \Big] + \frac{1}{4} \sum_{\otwo \neq \ol} \frac{\oone\otwo(\ol + 2n)}{\ol + \otwo} \left( X_{21i\ell} - X_{\ell i 12} \right) \nonumber \\
& + \frac{1}{4} \sum_{\oone \neq \ol} \frac{\oone\otwo(\ol + 2n)}{\ol + \oone} \left( X_{12i\ell} - X_{\ell i 12} \right).
\end{align}

When $m^2 = 0$, we have contributions from
\begin{align}
\label{R3}
\overline{R}_{i\ell}^{(-+)} &=  \frac{1}{4} \sum_{\otwo \neq \ol} \frac{\otwo}{\ol - \otwo} Z^+_{i12\ell} + \frac{1}{4} \sum_{\oone \neq \ol} \frac{\oone}{\ol - \oone} Z^+_{i21\ell} + \frac{1}{4} \sum_{i \neq \ell} \left( \frac{2n - \ol}{2n} \right) Z^-_{12i\ell} \nonumber \\
& \quad + \frac{1}{4} \sum_{\oone \neq \oi} \frac{1}{\oi - \oone} \Big[ \oone \left( H_{i12\ell} - 2\otwo^2 X_{i12\ell} \right) - (2n - \ol) \left( H_{1i2\ell} - 2\otwo^2 X_{1i2\ell} \right) \Big] \nonumber \\
& \quad + \frac{1}{4} \sum_{\otwo \neq \oi} \frac{1}{\oi - \otwo} \Big[ \otwo \left( H_{i21\ell} - 2\oone^2 X_{i21\ell} \right) - (2n - \ol) \left( H_{2i1\ell} - 2\oone^2 X_{2i1\ell} \right) \Big] \nonumber \\
& \quad - \frac{1}{8n} \sum_{\oone \neq \otwo} \Big[ \oone H_{21i\ell} + \otwo H_{12i\ell} - 2 \left( 2n - \ol \right)^2 \left(\oone X_{21i\ell} + \otwo X_{12i\ell} \right) \Big] \nonumber \\
& \quad - \frac{1}{2} \Big[ (2n - \ol) \left( \oone X_{21i\ell} + \otwo X_{12i\ell} \right) - \oone \otwo X_{i12\ell} \Big] .
\end{align}
{\it NB.}\, In \eqref{R3} only, $\oi = 2i + \Delta^+ = 2i + d$ since this term requires that $m^2 = 0$ to contribute. We maintain the same notation out of convenience, despite the special case. Finally, 
\begin{align}
\label{R1}
\overline{U}^{(1)}_{i\ell} &= \sum_{\oi \neq \oone} \left[ \frac{\oone^2}{\oi^2 - \oone^2} \left( H_{i11\ell} + m^2 V_{11i\ell} -2\oone^2 X_{i11\ell} \right) - \frac{\oi^2}{\oi^2 - \oone^2} \left( H_{1i1\ell} +m^2 V_{i11\ell} -2\oone^2 X_{11i\ell} \right) \right] \nonumber \\
& \!\!\!\! + \frac{1}{2} \sum_{\ol \neq \oone} \left[ \frac{\oone}{\ol + \oone} Z^-_{i11\ell} + \frac{\oone}{\ol - \oone}Z^+_{i11\ell} \right] + \frac{1}{2} \ol^2 \tilde{Z}^+_{11\ell} - \oone^2 X_{i11\ell} - \frac{1}{2}m^2 V_{i11\ell}  \, ,
\end{align}
and
\begin{align}
	\label{T1}
	\overline{T}^{(1)}_\ell = -\frac{1}{2} \left[ H_{11\ell\ell} + m^2 V_{\ell 11\ell} - 2\ol^2 X_{11\ell\ell} +4 \ol^2 \oone^2 P_{\ell\ell 1} + 2\oone^2 M_{\ell\ell 1} + 2m^2 \oone^2 Q_{\ell\ell 1} \right] \, .
\end{align}
Where $\overline{U}^{(2)}_{i\ell}$ and $\overline{T}^{(2)}_{\ell}$ follow from \eqref{R1} and \eqref{T1} by replacing $\oone \to \otwo$.

\begin{figure}
\centering
\includegraphics[width=\textwidth]{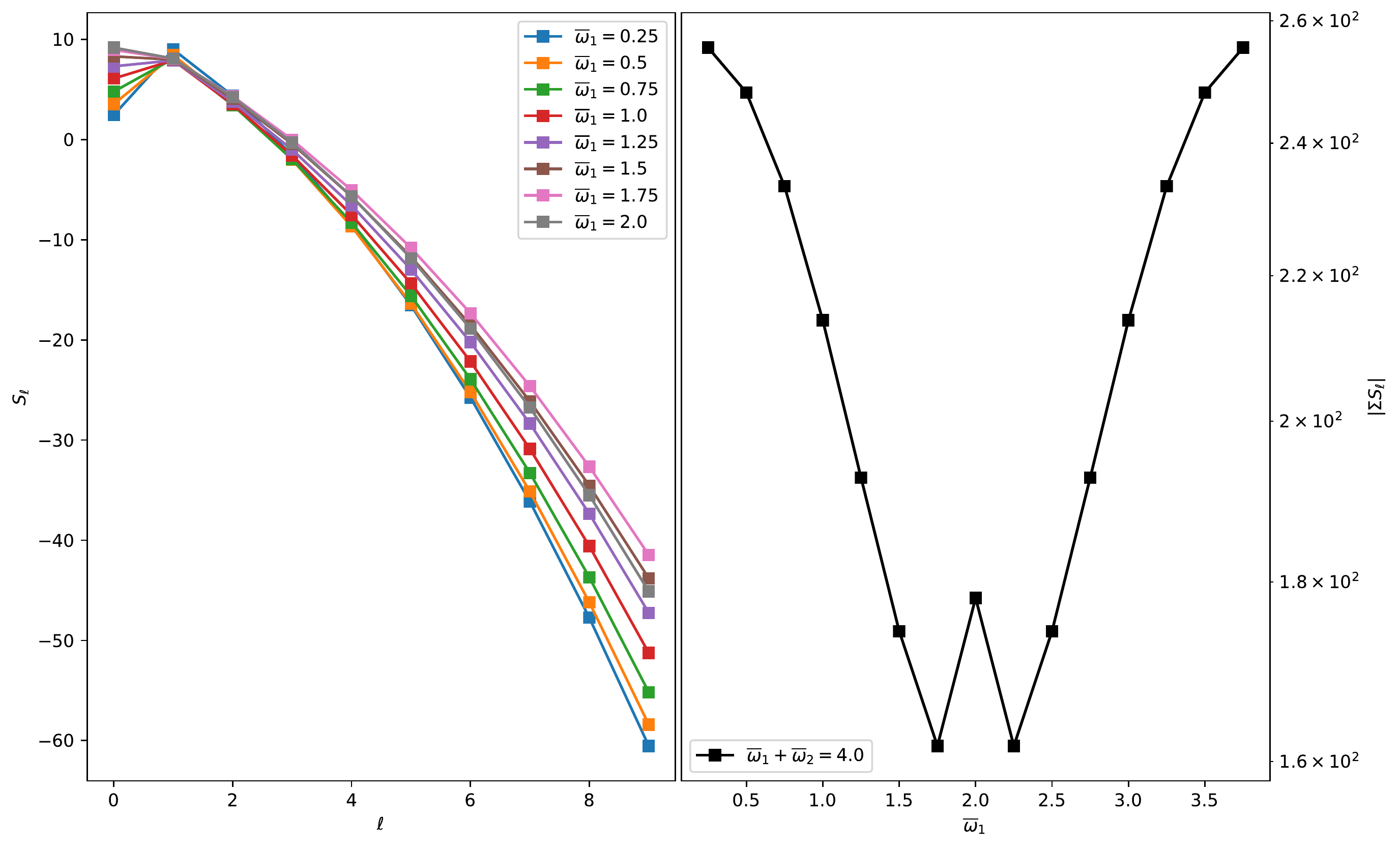}
\caption{{\it Left}: Source term values for a tachyonic scalar with $m^2 = -1.0$ when the frequencies of non-normalizable modes sum to $4.0$. {\it Right}: The absolute value of the sum of the source terms for each choice of $\oone$, $\otwo$.}
\label{fig:atoi_all_m-4_0}
\end{figure}

\begin{figure}[h!]
\centering
	\begin{subfigure}[b]{0.9\textwidth}
		\includegraphics[width=\textwidth]{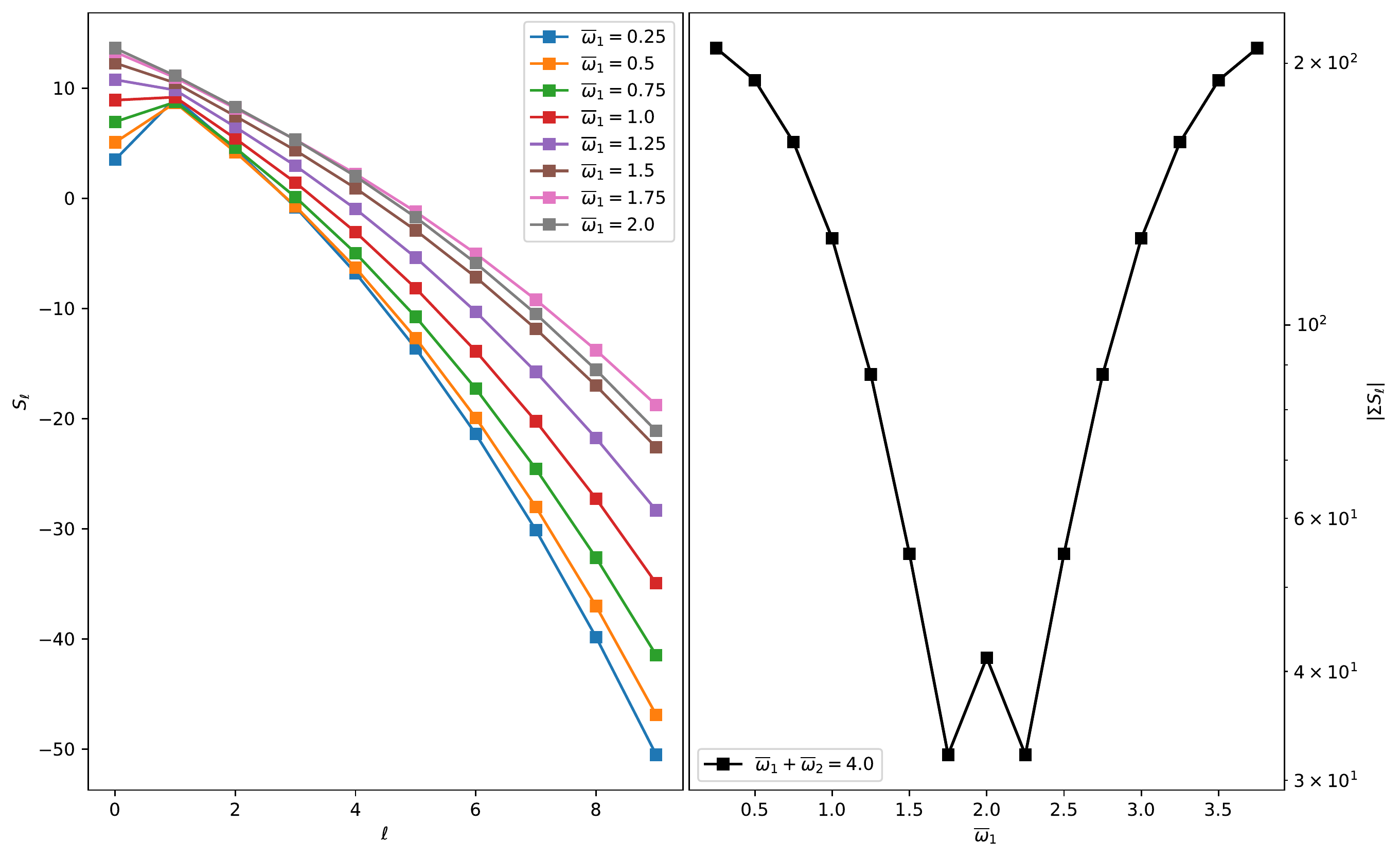}
		\label{fig:atoi_all_n2_m0}
	\end{subfigure}
	\vspace{-0.25in}
	\begin{subfigure}[b]{0.9\textwidth}
		\includegraphics[width=\textwidth]{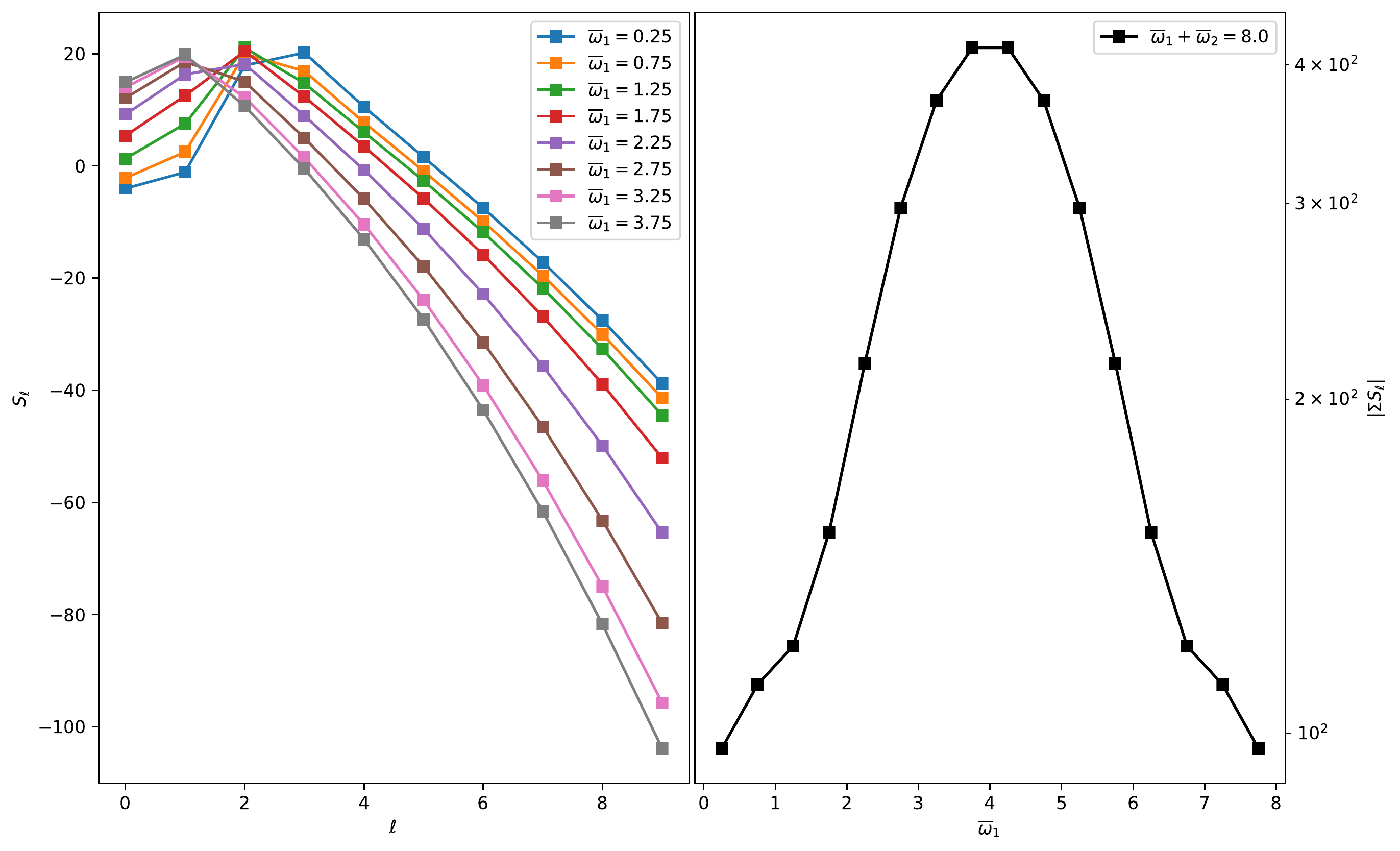}
		\label{fig:atoi_all_n4_m0}
	\end{subfigure}
	\caption{Evaluating \eqref{add to integer} for $m^2 = 0$ in $d=4$. {\it Above:} $S_\ell$ as a function of $\ell$ for a massless scalar with values of $\oone$ and $\otwo$ chosen so that $\oone + \otwo = 4$. {\it Below:} The same plot but with values chosen to satisfy $\oone + \otwo = 8$ so that contributions from $\overline R^{(-+)}_{i\ell}$ are present.}
	\label{fig:atoi_all_m0_0compare}
\end{figure}

In figure~\ref{fig:atoi_all_m-4_0}, we compute the total source term (modulo the amplitudes $a_i$ and $\bar A_\alpha$) for a tachyonic scalar with $n = 2$. Figure~\ref{fig:atoi_all_m0_0compare} provides a comparison between the value of the source term for a massless scalar between two choices of $n$: one that includes contributions from $\overline{R}_{i\ell}^{(-+)}$ and one that does not. As expected, the source terms are symmetric in $\oone \leftrightarrow \otwo$, hence only $\oone \leq n$ data are shown. Evaluations of each channel separately found that none vanished naturally. Since the total source term $S_\ell$ given by \eqref{add to integer} initially evaluates to small, positive values before becoming increasingly negative as $\ell$ becomes large, one may ask if the \emph{sum} of the channels vanishes. As a check for this, the absolute value of the sum of $S_\ell$ is also plotted; however, there is no indication that any channel vanishes for any of the $\oone$, $\otwo$ values considered. 

The renormalization flow equations include the sum of all the channels (none of which vanish naturally), and are
\begin{align}
\label{atoi RN 1}	
\frac{2 \ol}{\epsilon^2} \frac{d a_\ell}{d t} &= - \!\!\!\! \sum_{\oone + \otwo = 2n}\bigg[ \Theta\left( n - \ell - d \right) \overline{R}^{(-+)}_{(n - \ell - d) \ell} \ \bar A_1 \bar A_2 \, a_{(n - \ell - d)} \sin \left( b_{(n - \ell - d)} - \bar B_1 - \bar B_2 \right) \bigg]_{m^2 = 0}  \nonumber \\ 
& \quad - \!\!\!\! \sum_{\oone + \otwo = 2n} \!\!\!\! \Theta \left( \ell - n \right)  \overline{R}^{(++)}_{(\ell - n)\ell} \, \bar A_1 \bar A_2 \, a_{(\ell - n)} \sin \left( b_{(\ell - n)} + \bar B_1 + \bar B_2 \right) \nonumber \\
& \quad  - \!\!\!\! \sum_{\oone + \otwo = 2n} \!\!\!\! \overline{R}^{(+-)}_{(\ell + n) \ell} \, \bar A_1 \bar A_2 \, a_{(\ell + n)} \sin \left( b_{(\ell + n)} - \bar B_1 - \bar B_2 \right) \, ,
\end{align}
and
\begin{align}
\label{atoi RN 2}	
\frac{2 \ol a_\ell}{\epsilon^2} \frac{d b_\ell}{d t} &= - \!\!\!\! \sum_{\oone + \otwo = 2n}\bigg[ \Theta\left( n - \ell - d \right) \overline{R}^{(-+)}_{(n - \ell - d) \ell} \ \bar A_1 \bar A_2 \, a_{(n - \ell - d)} \cos \left( b_{(n - \ell - d)} - \bar B_1 - \bar B_2 \right) \bigg]_{m^2 = 0}  \nonumber \\ 
& \quad - \!\!\!\! \sum_{\oone + \otwo = 2n} \!\!\!\! \Theta \left( \ell - n \right)  \overline{R}^{(++)}_{(\ell - n)\ell} \, \bar A_1 \bar A_2 \, a_{(\ell - n)} \cos \left( b_{(\ell - n)} + \bar B_1 + \bar B_2 \right) \nonumber \\
& \quad  - \!\!\!\! \sum_{\oone + \otwo = 2n} \!\!\!\! \overline{R}^{(+-)}_{(\ell + n) \ell} \, \bar A_1 \bar A_2 \, a_{(\ell + n)} \cos \left( b_{(\ell + n)} - \bar B_1 - \bar B_2 \right) \nonumber \\
& \quad - \!\!\!\! \sum_{\oone + \otwo = 2n} \!\!\!\! a_\ell \left( \bar A_1^2 ( \overline{U}^{(1)}_{(\ell)\ell} + \overline{T}^{(1)}_\ell) + \bar A_2^2 (\overline{U}^{(2)}_{(\ell)\ell} + \overline{T}^{(2)}_\ell) \right) \, .
\end{align}


\subsection{Integer Plus $\chi$}
\label{ssec: intpluschi}

Finally, let us consider the case where the boundary condition can be written in terms of a sum of non-normalizable modes with constant amplitudes $\bar A_\gamma$ and frequencies $\omega_\gamma$ that are each non-integer valued, but differ from integer values by a set amount, i.e.
\begin{align}
	\label{int plus chi}
	\mc F(t) = \sum_{\gamma} \bar A_{\gamma} \cos \left( \omega_\gamma t + \bar B_{\gamma} \right) \quad \text{with} \quad \ogam = 2\gamma + \chi \, ,
\end{align}
where $\gamma \in \mathbb{Z}^+$. Greek letters are chosen to differentiate these non-normalizable modes from normalizable modes with integer frequencies, which use Roman letters. We furthermore limit $\chi$ to be non-integer\footnote{Indeed, for integer values of $\chi$, the sum or difference of two non-normalizable modes could be an integer. This would either be covered by the work in \S\!~\ref{ssec: add to integer}, or be a slight variation of it.} and set $m^2 = 0$ throughout. For this choice of non-normalizable frequencies there are no resonant contributions from the all-plus channel, unlike the naturally vanishing resonance found in \S\!~\ref{ssec: zero resonance}. Only when either $\oi + \ogam = \obet - \ol$, or $\oi + \ogam = \obet + \ol$ with $i + \gamma \geq \ell$, are resonant terms present. Let us examine each case separately.


\subsubsection{$\oi + \ogam = \obet - \ol$}
\label{sssec: intpluschi1}

The resonance condition $\oi + \ogam = \obet - \ol$ will contribute two types of terms to $S_\ell$. The first term includes a double sum over the frequencies $\omega_i$ and $\omega_\gamma$ with $\beta = i + \gamma + \ell$ and $\beta \neq \gamma$, and the second term includes a single sum over $\omega_\gamma$ with $\beta = \gamma + 2\ell + d$. In total, the source term is of the form
\begin{align}
\label{intpluschi1 source}
S_\ell &= \sum_{i \neq \ell} \sum_{\gamma \neq \beta} \overline{S}^{(1)}_{i (i + \gamma + \ell) \gamma \ell} \, a_i \bar A_{(i + \gamma + \ell)} \bar A_\gamma \cos \left( \theta_i - \theta_{(i + \gamma + \ell)} + \theta_\gamma \right) \nonumber \\
& \qquad + \sum_\gamma \overline{R}^{(1)}_{(\gamma + 2\ell + d) \ell} \, a_\ell \bar A_\gamma \bar A_{(\gamma + 2\ell + d)}  \cos \left(\theta_\ell - \theta_{(\gamma + 2\ell + d)} + \theta_\gamma \right)  \, ,
\end{align}
where 
\begin{align}
\overline{S}^{(1)}_{i \beta\gamma\ell} &= \frac{1}{4} H_{\beta\gamma i \ell} \frac{ \ogam (\oi - \obet + 2\ogam)}{(\obet - \ogam)(\oi + \ogam)} - \frac{1}{4} H_{\gamma\beta i \ell} \frac{\obet(\oi + \ogam - 2\obet)}{(\oi - \obet)(\obet - \ogam)} - \frac{1}{4} H_{\gamma i \beta\ell} \frac{\oi (\ogam - \obet + 2\oi)}{(\oi - \obet)(\oi + \ogam)} \nonumber \\
& + \frac{1}{2} \oi \ogam X_{\beta\gamma i \ell} \left( \frac{\ogam}{\oi - \obet} - \frac{\oi}{\obet + \ogam} + 1 \right) + \frac{1}{2} \oi \obet X_{\gamma\beta i \ell} \left( \frac{\oi}{\obet - \ogam} + \frac{\obet}{\oi + \ogam} - 1 \right) \nonumber \\
& + \frac{1}{2} \obet \ogam X_{i\beta\gamma\ell} \left(\frac{\obet}{\oi + \ogam} - \frac{\ogam}{\oi - \obet} - 1 \right) - \frac{1}{4} Z^+_{\beta\gamma i \ell} \left( \frac{\oi}{\oi + \ol}\right)  \nonumber \\
& + \frac{1}{4} Z^{-}_{i\gamma\beta\ell} \left(  \frac{\obet}{\ol - \obet}  \right) + \frac{1}{4} Z^+_{i\beta\gamma\ell} \left( \frac{\ogam}{\ol + \ogam} \right) \, ,
\end{align}
and
\begin{align}
\overline{R}^{(1)}_{\beta \ell} &= -\frac{1}{4} Z^{-}_{\gamma \ell \beta \ell} \left( \frac{\obet}{\ol - \obet} \right) -  \frac{1}{4} Z^+_{\ell \beta \gamma \ell} \left(\frac{\ogam}{\ol + \ogam} \right) + \frac{1}{8} H_{\beta\gamma\ell\ell} \frac{\ogam (\ogam - \ol)}{\ol (\ol + \ogam)} + \frac{1}{8} H_{\gamma\beta\ell\ell} \frac{\obet (\ol - \ogam)}{\ol (\ol + \ogam)} \nonumber \\
& - \frac{1}{4}\ogam\ol X_{\beta\gamma\ell\ell}\frac{(\ol - \ogam)}{(\ol + \ogam)} + \frac{1}{4} \obet \ol X_{\gamma\beta\ell\ell} \frac{(\ol + \ogam + 2\obet)}{(\ol + \ogam)} + \frac{1}{2} \obet \ogam X_{\ell\ell\beta\gamma} \frac{(\ogam + \obet)}{(\ol + \ogam)} \nonumber \\
& -\frac{1}{2} \ol \obet X_{\gamma\beta\ell\ell} - \frac{1}{2}\obet \ogam X_{\ell\ell\beta\gamma} + \frac{1}{2} \ol \ogam X_{\beta\gamma\ell\ell} - \ol^2 \tilde{Z}^+_{\beta\gamma\ell} \, .
\end{align}


\subsubsection{$\oi + \ogam = \obet + \ol$}
\label{ssec: intpluschi2}

Similarly, when the resonance condition $\oi + \ogam = \obet + \ol$ is met, the contribution to the source term is
\begin{align}
\label{intpluschi2 source}
S_\ell &= \underbrace{\sum_{i \neq \ell} \sum_{\gamma \neq \beta}}_{i + \gamma \geq \ell} \overline{S}^{(2)}_{i (i + \gamma - \ell) \gamma \ell} \, a_i \bar A_{(i + \gamma - \ell)} \bar A_\gamma \cos \left( \theta_i - \theta_{(i + \gamma - \ell)}  + \theta_\gamma \right) \nonumber \\
& \qquad \qquad + \sum_\beta \overline{R}^{(2)}_{\beta\ell} \, a_\ell \bar A_\beta^2 \cos \left( \theta_\ell + \theta_\beta - \theta_\beta \right) \, ,
\end{align}
where
\begin{align}
\overline{S}^{(2)}_{i\beta\gamma\ell} &= \frac{1}{4} H_{\beta\gamma i\ell} \frac{\ogam (\oi - \obet)}{(\obet - \ogam)(\oi - \ogam)} - \frac{1}{4} H_{\gamma\beta i \ell} \frac{\obet(\ol - \obet)}{(\obet - \ogam)(\oi - \obet)} + \frac{1}{4} H_{\beta i \gamma\ell} \frac{\oi (\ogam - \obet)}{(\oi - \obet)(\oi - \ogam)} \nonumber \\
& + \frac{1}{2} \oi \ogam X_{\beta\gamma i \ell} \left( \frac{\ogam}{\oi - \obet} - \frac{\oi}{\obet - \ogam} + 1 \right) + \frac{1}{2} \oi \obet X_{\gamma\beta i \ell} \left( \frac{\oi}{\obet - \ogam} - \frac{\obet}{\oi - \ogam} - 1 \right) \nonumber \\
& + \frac{1}{2} \obet \ogam X_{i\beta\gamma\ell} \left( \frac{\obet}{\oi - \ogam} - \frac{\ogam}{\oi - \obet} - 1 \right)  + \frac{1}{4} Z^-_{i\gamma\beta\ell} \left( \frac{\obet}{\ol + \obet}\right)  \nonumber \\
&
+ \frac{1}{4} Z^+_{i\beta\gamma\ell} \left( \frac{\ogam}{\ol - \ogam}\right) - \frac{1}{4}Z^+_{\beta\gamma i \ell} \left( \frac{\oi}{\oi - \ol} \right) \, ,
\end{align}
and
\begin{align}
\overline{R}^{(2)}_{\beta\ell} &= \frac{1}{4} Z^-_{\ell\beta\beta\ell} \left( \frac{\obet}{\ol + \obet} \right) + \frac{1}{4} Z^+_{\ell\beta\beta\ell} \left( \frac{\obet}{\ol - \obet} \right) + \frac{1}{2} H_{\ell\beta\beta\ell} \left( \frac{\obet^2}{\ol^2 - \obet^2} \right) - \frac{1}{2} H_{\beta\ell\beta\ell} \left( \frac{\ol^2}{\ol^2 - \obet^2} \right) \nonumber \\
& \!\!\!\!\!\!\!\! + X_{\beta\beta\ell\ell} \left( \frac{\ol^2}{\ol^2 - \obet^2} \right) + \frac{1}{2} \obet^2 X_{\ell\beta\beta\ell} \left( \frac{\ol^2 + \obet^2}{\ol^2 - \obet^2} \right) - \frac{1}{2} H_{\beta\beta\ell\ell} +  \ol^2 \tilde{Z}^+_{\beta\beta\ell} - 2 \obet^2 \ol^2 P_{\ell\ell\beta} - \obet^2 M_{\ell\ell\beta} . \vspace{-0.3in}
\end{align}

\begin{figure}[t]
\centering
	\includegraphics[width=\textwidth]{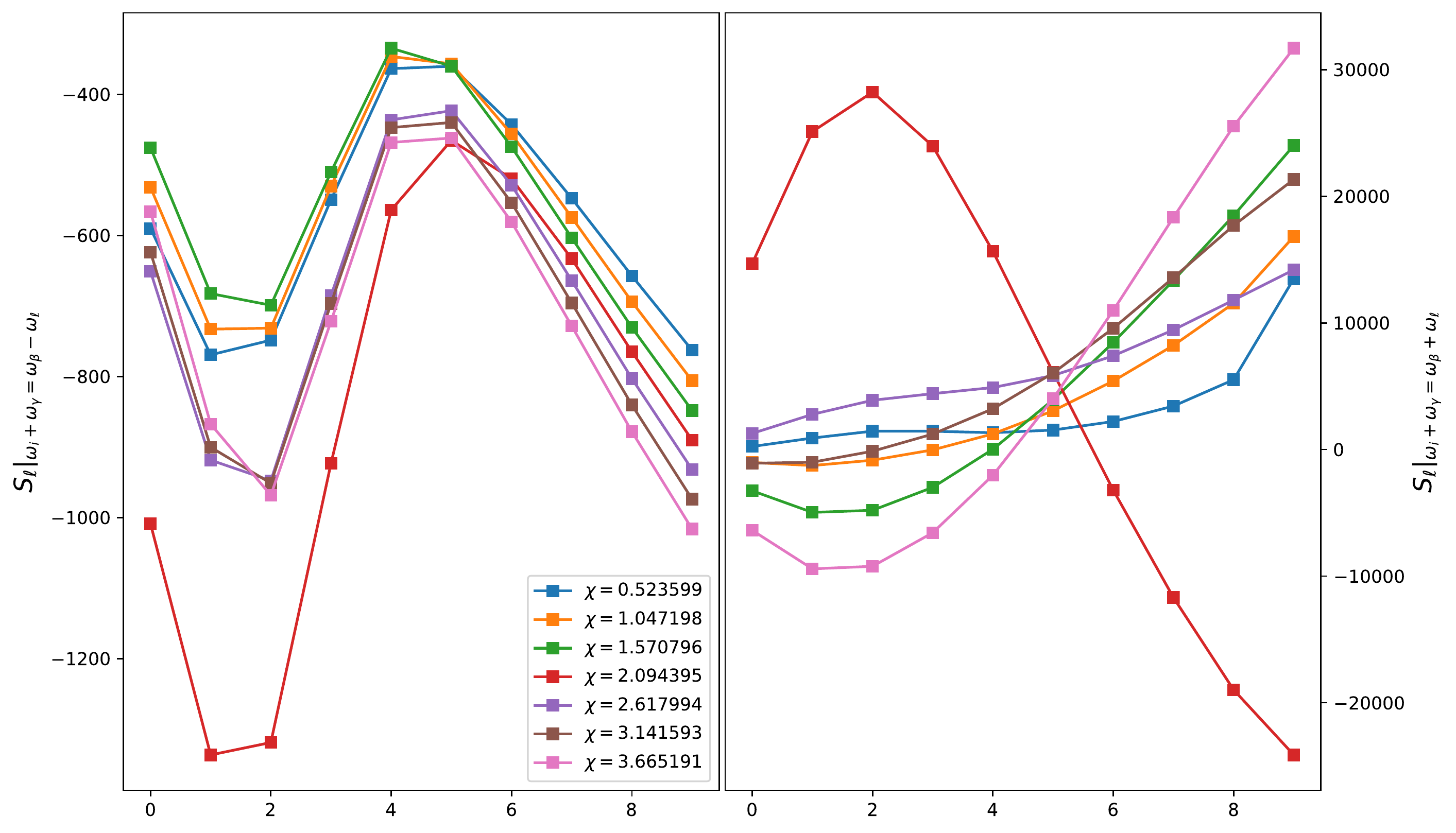}
	\caption{{\it Left:} Evaluating the source term \eqref{intpluschi1 source} for various values of $\chi$ for $\ell < 10$. {\it Right:} Evaluating the source term \eqref{intpluschi2 source} subject to $i + \gamma \geq \ell$ for the same values of $\chi$ and the same range of $\ell$.}
	\label{fig: twoiplusx}
\end{figure}

Unlike the case with all normalizable modes where two of the three resonance channels naturally vanished, both of the resonant channels contribute when the non-normalizable modes have frequencies given by \eqref{int plus chi}. Therefore, the renormalization flow equations will contain contributions from both channels:
\begin{align}
\frac{2 \ol}{\epsilon^2} \frac{d a_\ell}{d t} &= - \sum_{i \neq \ell} \sum_{\gamma \neq \beta} \overline{S}^{(1)}_{i (i + \gamma + \ell) \gamma \ell} \, a_i \bar A_{(i + \gamma + \ell)} \bar A_\gamma \sin \left( b_\ell + \bar B_{(i + \gamma + \ell)} - b_i - \bar B_\gamma \right) \nonumber \\
& \qquad \qquad -  \underbrace{\sum_{i \neq \ell} \sum_{\gamma \neq \beta}}_{i + \gamma \geq \ell} \overline{S}^{(2)}_{i (i + \gamma - \ell) \gamma \ell} \, a_i \bar A_{(i + \gamma - \ell)} \bar A_\gamma \sin \left( b_\ell + \bar B_{(i+\gamma - \ell)} - b_i - \bar B_\gamma \right) \, ,
\end{align}
\begin{align}
\frac{2 \ol a_\ell}{\epsilon^2} \frac{d b_\ell}{dt} &=  - \sum_\beta \overline{R}^{(2)}_{\beta\ell} \, a_\ell \bar A_\beta^2 - \sum_\gamma \overline{R}^{(1)}_{(\gamma + 2\ell + d) \ell} \, a_\ell \bar A_\gamma \bar A_{(\gamma + 2\ell + d)} \cos \left(b_\ell - \bar B_{(\gamma + 2\ell + d)} + \bar B_\gamma  \right) \nonumber \\
& \qquad - \sum_{i \neq \ell} \sum_{\gamma \neq \beta} \overline{S}^{(1)}_{i (i + \gamma + \ell) \gamma \ell} \, a_i \bar A_{(i + \gamma + \ell)} \bar A_\gamma \cos \left( - b_\ell + \bar B_{(i + \gamma + \ell)} - b_i - \bar B_\gamma \right) \nonumber \\
& \qquad \qquad -  \underbrace{\sum_{i \neq \ell} \sum_{\gamma \neq \beta}}_{i + \gamma \geq \ell} \overline{S}^{(2)}_{i (i + \gamma - \ell) \gamma \ell} \, a_i \bar A_{(i + \gamma - \ell)} \bar A_\gamma \cos \left( b_\ell + \bar B_{(i+\gamma - \ell)} - b_i - \bar B_\gamma \right) \, .
\end{align}
In figure~\ref{fig: twoiplusx}, we evaluate both resonant contributions channels' and plot their contributions for various values of $\chi$. In particular, we examine the values $\chi \in \{ \pi/6, \ldots, 7\pi/6 \}$. Again, there is no indication of any channel vanishing naturally. Interestingly, both sources demonstrate anomalous behaviour when $\chi \sim 2$ for reasons that are not immediately clear. The source term \eqref{intpluschi1 source} is generally more positive for larger $\chi$ except for $\chi = 2 \pi / 3$, which is translated negatively with respect to the source terms produced by other $\chi$ values. Again, when \eqref{intpluschi2 source} is evaluated for $\chi = 2 \pi/3$, the result differs significantly from other choices of $\chi$: seemingly reflected through the $x$~axis with respect to other results. The significance of the choice $\chi = 2\pi / 3 \sim d / 2$ is possibly explained by the non-normalizable modes being \emph{nearly} equal to the normalizable ones. In this event, $S_\ell$ would contain additional terms, such as those present in \S~\!\ref{sec: norm res}. The departure of the $\chi = 2\pi/3$ data from other data sets is perhaps a signal of these missing resonances.


\section{Discussion}
\label{sec: discussion}

We have seen that the inclusion of a time-dependent boundary term in the holographic dual of a quantum quench allows energy to enter the bulk spacetime through coupling with non-normalizable modes. The dynamics of the weakly turbulent energy cascades that trigger instability were captured by secular terms at third-order that could not be removed by frequency shifts alone. Using the Two-Time Formalism, we have determined the flow equations for the slowly varying amplitudes and phases that are tuned to cancel the secular terms at each instant. 

Following our discussion at the end of \S\!~\ref{sec: source terms and BCs}, we have limited the types of resonances we have considered to only those produced when the resonance condition ${\omega_I \pm \omega_J \pm \omega_K = \pm \omega_\ell}$ contains \emph{exactly two} non-normalizable frequencies. As a result, the flow equations for the time-dependent amplitudes and phases of the normal modes are linear in the amplitude. When only normalizable modes are considered, these equations contain three powers of the amplitude~\cite{1403.6471, 1407.6273}. It is important to note that the resonances from only normalizable modes are still present (see appendix~\ref{sec: norm res} for their explicit forms) and would also contribute to the non-normalizable resonances. Furthermore, we have restricted our choice of mass to within $m_{BF}^2 < m^2 \leq 0$ to avoid potential issues raised by renormalizability of the metric functions ${A(t,x),~\delta(t,x)}$ on the boundary of AdS. Future work will consider what counterterms are required to keep the CFT renormalizable outside of this restriction on the mass. What remains to be shown is if, for some judicious choice of time-dependent boundary term, the different types of resonances would destructively interfere and render the perturbative description naturally stable. 

We have considered a set of broadly applicable choices for the form of the driving term $\mc F(t)$ that require little or no fine tuning to produce resonant contributions. When the driving term is given by a single, non-integer frequency component, there is a single resonant channel and the presence of non-normalizable modes affects only the flow of the phase. While both $a_\ell$ and $b_\ell$ will also receive terms due to all-normalizable resonances, the fact that this choice of driving term induces extra drift in the phase only is interesting. By considering more specialized forms of the driving term, additional resonance channels from non-normalizable modes are present. Unlike when only normalizable modes are considered, none of the channels are naturally vanishing. When the driving term is given by the superposition of two periodic functions with frequencies that add to an integer, the flow equations for $a_\ell$ and $b_\ell$ receive contributions from multiple terms that mix modes. For example, \eqref{atoi RN 1} shows that $da_\ell / dt$ can contain contributions proportional to $a_{(n-\ell-d)}$, $a_{(\ell -n)}$, and $a_{(\ell + n)}$. Further mixing of modes is observed when the driving term is given by a Fourier sum over frequencies that differ from an integer by a set amount. In this case, the equations for $da_\ell / dt$ and $db_\ell / dt$ both receive contributions from mixed normalizable amplitudes and non-normalizable amplitudes, e.g.~$da_\ell / dt$ contains terms proportional to both ${\bar A_{(i + \gamma +\ell)}}$ and ${\bar A_{(i + \gamma - \ell)}}$. Unlike when only normalizable modes are considered, there were no naturally vanishing resonance channels for the frequencies considered.

With the renormalization flow equations established, future work will compare the perturbatively stable solutions found from the TTF theory to the behaviour of the fully numerical systems in the limit of weak driving~\cite{1712.07637, 1502.05726}. Comparisons to established numerical pumped solutions in the full theory may be instructive in understanding the space of stable and nearly-stable data. Indeed, this was the case with the TTF description of a collapsing scalar field subject static boundary conditions~\cite{1507.08261}. Furthermore, the development of a perturbative description for pumped scalars may aide in solving the fully nonlinear system by reducing computational overhead when the full system is known to be well described by the TTF. Beyond this time, numerical methods could switch to conventional, computationally-complex methods as the system approaches collapse.


\acknowledgments The author would like to thank A. R. Frey for their guidance and insight with this project. This work is supported by the Natural Sciences and Engineering Research Council of Canada's Discovery Grant program.


\appendix
\section{Derivation of Source Terms For Massive Scalars}
\label{app: source term derivation}

The derivation of the general expression for the $\mc{O}(\epsilon^3)$ source term for massive scalars closely follows the massless case, particularly if one chooses not to write out the explicit mass dependence as was done in \cite{1810.04753}. However, since we have chosen to write our equations in a slightly different way -- and in a different gauge -- than previous authors, one may find it instructive to see the differences in the derivations. Below we have included the intermediate steps involved in deriving the third-order source term $S_\ell$.

Continuing the expansion of the equations of motion in powers of $\epsilon$, we see that the backreaction between the metric and the scalar field appears at second order in the perturbation,
\begin{align}
A_2' = - \mu \nu \left[ (\dot \phi_1 )^2 + (\phi_1')^2 + m^2 \phi_1^2 \sec^2 x \right] + \nu' A_2 / \nu \, ,
\end{align}
which can be directly integrated to give
\begin{align}
A_2 = -\nu \int^x_0 dy \, \mu \left( (\dot \phi_1 )^2 + (\phi_1')^2 + m^2 \phi_1^2 \sec^2 x \right) \, .
\end{align}
For convenience, we have also defined the functions
\begin{align}
\mu (x) = \left( \tan x \right)^{d-1} \quad \text{and} \quad \nu(x) = (d-1) / \mu ' \, .
\end{align}
Similarly, the first non-trivial contribution to the lapse (in the boundary time gauge) is
\begin{align}
\delta_2 = \int^{\pi/2}_x dy \, \mu \nu \left(  (\dot \phi_1 )^2 + (\phi_1')^2 \right) \, .
\end{align}

Projecting each of the terms in \eqref{3rd order} individually onto the eigenbasis $\{ e_\ell \}$ will involve evaluating inner products involving multiple integrals. To aide in evaluating these expressions, it is useful to derive several identities. First, from the equation for the scalar field's time-dependent coefficients $c_i$,
\begin{align} 
\ddot c_i + \oi^2 c_i = 0 \quad \Rightarrow \quad \p_t \left(\dot c_i^2 + \oi^2 c_i^2 \right) = \p_t \mathbb C_i = 0 \, .
\end{align}
Next, from the definition of $\hat L$,
\begin{align}
\hat L e_j = -\frac{1}{\mu} \left( \mu e'_j \right)' + m^2 \sec^2 x e_j \quad \Rightarrow \quad \left( \mu e'_j \right)' = \mu \left( m^2 \sec^2 x - \omega_j^2 \right) e_j \, .
\end{align}
By considering the expression $\left( \mu e'_i e_j \right)'$, we see that
\begin{align}
\left( \mu e'_i e_j \right) ' = \left(m^2 \sec^2 x - \oi^2 \right) \mu e_i e_j + \mu e'_i e'_j \, ,
\end{align}
which, after permuting $i, j$ and subtracting from above, gives
\begin{align}
\frac{\left[ \mu (e'_i e_j \oj^2 - e_i e'_j \oi^2 ) \right]'}{(\oj^2 - \oi^2)} = \mu m^2 \sec^2 x e_i e_j + \mu e'_i e'_j \, .
\end{align}

Using these identities, we evaluate each of the inner products and find that
\begin{align}
\label{inner prod 1}
\langle \delta_2 \ddot \phi_1, e_\ell \rangle &= - \sum_{i = 0}^\infty \sum_{\substack{j=0 \\ k \neq \ell}}^\infty \sum_{k=0}^\infty \frac{\ok^2 c_k}{\ol^2 - \ok^2} \left[\dot c_i \dot c_j \left(X_{k\ell ij} - X_{\ell k i j} \right) + c_i c_j \left( Y_{ij\ell k} - Y_{ijk\ell} \right) \right] \nonumber \\
& \qquad  - \sum_{i=0}^\infty \sum_{j=0}^\infty \ol^2 c_\ell \left[ \dot c_i \dot c_j P_{ij\ell} + c_i c_j B_{i j \ell} \right] \, , \\
\langle A_2 \ddot \phi_1, e_\ell \rangle &= 2 \sum_{i = 0}^\infty \sum_{\substack{j=0 \\ i \neq j}}^\infty \sum_{k=0}^\infty \frac{\ok^2 c_k}{\oj^2 - \oi^2} X_{ijk \ell} \left( \dot c_i \dot c_j + \oj^2 c_i c_j \right) \nonumber \\
& \qquad + \sum_{i = 0}^\infty \sum_{j = 0}^\infty \oj^2 c_j \left( \mathbb C_i P_{j \ell i} + c_i^2 X_{ii j \ell} \right) \, , \\
\langle \dot \delta_2 \dot \phi_1 , e_\ell \rangle &= \sum_{i = 0}^\infty \sum_{\substack{j=0 \\ k \neq \ell}}^\infty \sum_{k=0}^\infty \frac{\dot c_k}{\ol^2 - \ok^2} \left[ \p_t \left( \dot c_i \dot c_j \right) \left( X_{k\ell ij} - X_{\ell k i j} \right) + \p_t (c_i c_j) \left(Y_{ij\ell k} - Y_{ijk\ell}\right) \right] \nonumber \\
& \qquad+ \sum_{i=0}^\infty \sum_{j=0}^\infty \dot c_\ell \left[ \p_t \left( \dot c_i \dot c_j \right) P_{ij\ell} + \p_t (c_i c_j) B_{ij\ell} \right] \, , \\
\langle \dot A_2 \dot \phi_1, e_\ell \rangle &= -2 \sum_{i=0}^\infty \sum_{j=0}^\infty \sum_{k=0}^\infty  \dot c_k \dot c_j c_i X_{ijk\ell} \, , \\
\langle \left( A_2' - \delta_2' \right) \phi_1', e_\ell \rangle &= - 2 \sum_{i = 0}^\infty \sum_{\substack{j=0 \\ i \neq j}}^\infty \sum_{k=0}^\infty \frac{c_k (\dot c_i \dot c_j + \oj^2 c_i c_j)}{\oj^2 -\oi^2} H_{ijk\ell} -m^2 \sum_{i=0}^\infty \sum _{j=0}^\infty \sum_{k=0}^\infty c_i c_j c_k V_{ijk\ell} \nonumber \\
& \qquad - \sum_{i=0}^\infty \sum_{j=0}^\infty c_j \left[ c_i^2 H_{iij\ell} + \mathbb C_i M_{j \ell i} \right] \, , \\
\label{inner prod 2}
\langle A_2 \phi_1 \sec^2 x, e_\ell \rangle &= - 2\sum_{i = 0}^\infty \sum_{\substack{j=0 \\ i \neq j}}^\infty \sum_{k=0}^\infty \frac{c_k (\dot c_i \dot c_j + \oj^2 c_i c_j )}{\oj^2 - \oi^2} V_{jki\ell} \nonumber \\
& \qquad - \sum_{i=0}^\infty \sum_{j=0}^\infty c_j \left( c_i^2 V_{jii\ell} + \mathbb C_i Q_{j\ell i} \right) ,
\end{align}
where the forms of X, Y, V, H, B, M, P, and Q are given by
\begin{align}
X_{ijk\ell} &= \int^{\pi/2}_0 dx \, \mu^2 \nu e'_i e_j e_k e_\ell \\
Y_{ijk\ell} &= \int^{\pi/2}_0 dx \, \mu^2 \nu e'_i e'_j e_k e'_\ell \\
V_{ijk\ell} &= \int^{\pi/2}_0 dx \, \mu^2 \nu e_i e_j e'_k e_\ell \sec^2 x \\
H_{ijk\ell} &= \int^{\pi/2}_0 dx \, \mu^2 \nu' e'_i e_j e'_k e_\ell \\
B_{ij\ell} &= \int^{\pi/2}_0 dx \, \mu \nu e'_i e'_j \int^x_0 dy \, \mu e^2_\ell \\
M_{ij\ell} &= \int^{\pi/2}_0 dx \, \mu \nu' e'_i e_j \int^x_o dy \, \mu e_\ell^2 \\
P_{ij\ell} &= \int^{\pi/2}_0 dx \, \mu \nu e_i e_j \int^x_0 dy \, \mu e^2_\ell \\
Q_{ij\ell} &= \int^{\pi/2}_0 dx \, \mu \nu e_i e_j \sec^2 x \int^x_0 dy \, \mu e^2_\ell \, .
\end{align}

Note that, using integration by parts to remove the derivative from $\nu$ in the definitions of $H_{ijk\ell}$ and $M_{ij\ell}$, we can show that
\begin{align}
H_{ijk\ell} &= \oi^2 X_{kij\ell} + \ok^2 X_{ijk\ell} - Y_{ij\ell k}  - Y_{\ell kji}   - m^2 V_{kji\ell} -m^2 V_{ijk\ell} \, , \\
M_{ij\ell} &= \oi^2 P_{ij\ell} - B_{ij\ell} -m^2 Q_{ij\ell} \, .
\end{align}

Collecting \eqref{inner prod 1}~\!-~\!\eqref{inner prod 2} gives the expression for $S_\ell = \langle S, e_\ell \rangle$:
\begin{align}
\label{S intermediate}
S_\ell &= \sum_{\substack{i, j, k \\ k \neq \ell}}^\infty \frac{1}{\ol^2 - \ok^2} \Big[ F_k(\dot c_i \dot c_j) \left(X_{k\ell i j} - X_{\ell k i j} \right) + F_k(c_i c_j) \left(Y_{ij\ell k} - Y_{ijk\ell} \right) \Big] \nonumber \\
& \quad +2 \sum_{\substack{i,j,k \\ i \neq j}}^\infty \frac{c_k D_{ij}}{\oj^2 - \oi^2} \Big[  2\ok^2 X_{ijk\ell} - H_{ijk\ell} -m^2 V_{jki\ell} \Big] - \sum_{i,j,k}^\infty c_i \Big[ 2 \dot c_j \dot c_k X_{ijk\ell} + m^2 c_j c_k V_{ijk\ell} \Big] \nonumber \\ 
& \quad + \sum_{i,j}^\infty \Big[ F_\ell (\dot c_i \dot c_j) P_{ij\ell} + F_\ell (c_i c_j) B_{ij\ell} + 2\oj^2 c_j \left( c_i^2 X_{iij\ell} + \mathbb C_i P_{j\ell i} \right) \nonumber \\
& \qquad - c_j \left( c^2_i (H_{iij\ell} + m^2 V_{jii\ell} ) + \mathbb C_i (M_{j\ell i} + m^2 Q_{j\ell i}) \right) \Big] \, ,
\end{align}
where $F_k(z) = \dot c_k \dot z - 2\ok^2 c_k z$, $D_{ij} = \dot c_i \dot c_j + \omega^2_j c_i c_j$, and $\mathbb C_i = \dot c_i^2 + \oi^2 c_i^2$. Additionally, we have combined some integrals into their own expressions, namely
\begin{align}
Z^{\pm}_{ijk\ell} = \oi \oj \left( X_{k\ell ij} - X_{\ell kij} \right) \pm \left( Y_{ij\ell k} - Y_{ijk\ell} \right) \quad \text{and} \quad \tilde Z^{\pm}_{ij\ell} = \oi \oj P_{ij\ell} \pm B_{ij\ell} \, .
\end{align}
Finally, using the solution for the time-dependent coefficients, $c_i(t) = a_i(t) \cos \left( \omega_i t + b_i (t) \right) \equiv a_i \cos \theta_i$, we arrive at \eqref{general source}.


\section{A Single Non-normalizable Mode: Integer Frequencies}
\label{more 2NN}
Let us return to the study of non-normalizable modes, and the particular case of the boundary term $\mc F(t)$ being given by a single function
\begin{align}
	\mc F(t) = \bar A_{\ob} \cos \ob t
\end{align}
with fixed amplitude $\bar A_{\ob}$ and some frequency $\ob$ that is not set \emph{a priori} to be integer or non-integer (recall also that $m^2_{BF} < m^2 \leq 0$ by the discussion in \S\!~\ref{sec: NNmodes}). In \S~\!\ref{ssec: equalNN} we considered values of $\ob$ that were strictly non-integer; however, within the space of resonant frequencies, there are are choices of $\ob \in \mathbb{Z}^+$ that may produce extra resonances. These instances were excluded from the discussion in \S\!~\ref{ssec: equalNN} and instead we address them here. When considering special integer values of $\ob$, each choice of $\ob$ below will contribute a $\overline T$-type term to the total source:
\begin{align}
\label{gen NN res 1}
\overline{T}^{(1)}_{i}: \quad \omega_i &= \ol + 2\ob \quad \forall \; \ob \in \mathbb{Z}^+ \\
\overline{T}^{(2)}_{i}: \quad \omega_i &= \ol - 2\ob \quad \forall \; \ob \in \mathbb{Z}^+ \; \text{such that } \ell \geq \ob \\
\label{gen NN res 2}
\overline{T}^{(3)}_{i}: \quad \omega_i &= 2\ob - \ol \quad \forall \; \ob \in \mathbb{Z}^+ \; \text{such that } \ob \leq \ell + \Delta^+ \, ,
\end{align}
with $\omega_i \neq \omega_\ell$ in each case. These special values contribute to the case of two, equal non-normalizable modes via
\begin{align}
\label{2NN all}
S_\ell &= \bar A^2_{\ob} \, \overline{T}^{(1)}_{(\ell + \ob)} \, a_{(\ell + \ob)} \cos \left( \theta_{(\ell + \ob)} - 2\ob t \right) + \bar A^2_{\ob} \, \overline{T}^{(2)}_{(\ell - \ob)} \, a_{(\ell - \ob)}\cos \left( \theta_{(\ell - \ob)} + 2\ob t \right) \nonumber \\
& \quad + \bar A^2_{\ob} \, \overline{T}^{(3)}_{(\ob - \ell - \Delta^+)} \, a_{(\ob - \ell- \Delta^+)} \cos \left( 2\ob t - \theta_{(\ob - \ell - \Delta^+)} \right) 
\end{align}
under their respective conditions on the value of $\ob$. The total resonant contribution for all possible $\ob$ values is the addition of \eqref{2NN all} and \eqref{2genNN}. 
Evaluating \eqref{general source} for each of the cases described by \eqref{gen NN res 1}~\!-~\!\eqref{gen NN res 2}, we find that
\begin{align}
\overline{T}^{(1)}_{i} &= \frac{1}{2} \bigg[ \, H_{i\ob\ob\ell} \left( \frac{\ob}{\oi - \ob} \right) - H_{\ob i \ob\ell} \left( \frac{\oi}{\oi - \ob} \right) + m^2 V_{\ob\ob i\ell} \left( \frac{\ob}{\oi - \ob} \right) \nonumber \\
& - m^2 V_{i \ob\ob\ell} \left( \frac{\oi}{\oi - \ob} \right) - 2 \ob^2 X_{i\ob\ob\ell} \left( \frac{\ob}{\oi - \ob} \right) + 2 \ob^2 X_{\ob i \ob\ell} \left( \frac{\oi}{\oi - \ob} \right) \bigg]_{\oi \neq \ob} \nonumber \\
& -\frac{1}{2} \bigg[ Z^+_{i\ob\ob\ell} \left( \frac{\ob}{\ol + \ob} \right) \bigg]_{\ol \neq \ob} \!\! + \frac{1}{4} Z^-_{\ob\ob i \ell} \left( \frac{\ol + 2\ob}{2 \ob} \right) + \frac{1}{2} \ob^2 X_{i\ob\ob\ell} - \frac{m^2}{4} V_{\ob\ob i \ell} \nonumber \\
& - \ob \oi X_{\ob\ob i\ell} - \frac{m^2}{2} V_{i \ob\ob\ell} \, ,
\end{align}
\begin{align}
\overline{T}^{(2)}_{i} &=  - \frac{1}{2} \bigg[ \, H_{i\ob\ob \ell} \left( \frac{\ob}{\oi + \ob} \right) + H_{\ob i \ob \ell} \left( \frac{\oi}{\oi + \ob} \right) + m^2 V_{\ob \ob i \ell} \left( \frac{\ob}{\oi + \ob} \right) \nonumber \\
& + m^2 V_{i\ob\ob\ell} \left( \frac{\oi}{\oi + \ob} \right) - 2 \ob^2 X_{i \ob\ob\ell} \left( \frac{\ob}{\oi + \ob} \right) - 2 \ob^2 X_{\ob\ob i\ell} \left( \frac{\oi}{\oi + \ob} \right) \bigg]_{\oi \neq \ob} \nonumber \\
& - \frac{1}{2} \bigg[ \, Z^{-}_{i \ob \ob \ell} \left( \frac{\ob}{\ol - \ob} \right) \bigg]_{\ol \neq \ob} \!\!\!\! - \frac{1}{4} Z^-_{\ob\ob i \ell} \left( \frac{\ol - 2\ob}{\ob} \right) + \frac{1}{2} \ob^2 X_{i\ob\ob\ell} + \frac{m^2}{4} V_{\ob\ob i\ell} \nonumber \\
& + \ob \oi X_{\ob\ob i \ell} + \frac{m^2}{2} V_{i \ob\ob \ell} \, ,
\end{align}
and
\begin{align}
\overline{T}^{(3)}_{i} &= \frac{1}{2} \bigg[ \, H_{i\ob\ob\ell} \left( \frac{\ob}{\oi - \ob} \right) - H_{\ob i \ob\ell} \left( \frac{\oi}{\oi - \ob} \right) + m^2 V_{\ob\ob i \ell} \left( \frac{\ob}{\oi - \ob} \right) \nonumber \\
& - m^2 V_{i\ob\ob\ell} \left( \frac{\oi}{\oi - \ob} \right) - 2 \ob^2 X_{i\ob\ob\ell} \left( \frac{\ob}{\oi-\ob} \right) + 2 \oi^2 X_{\ob\ob i\ell} \left( \frac{\ob}{\oi-\ob} \right) \nonumber \\
& - Z^+_{i\ob\ob\ell} \left( \frac{\ob}{\oi - \ob} \right) \bigg]_{\oi \neq \ob} + \frac{1}{4} Z^-_{\ob\ob i \ell} \left( \frac{2\ob - \ol}{2\ob} \right) + \frac{1}{2} \ob^2 X_{i\ob\ob\ell} - \frac{m^2}{4} V_{\ob\ob i\ell} \nonumber \\
&  - \ob \oi X_{\ob\ob i\ell} - \frac{m^2}{2} V_{i\ob\ob\ell} \, .
\end{align}
These resonance channels can then be added into the right hand side of the equation for $d a_\ell / d t$ in \eqref{equal NN flow}.


\section{Resonances From Normalizable Solutions}
\label{sec: norm res}
In this appendix, we include explicit expressions for contributions to the source term $S_\ell$ from a massive scalar with static boundary conditions ($\mc F(t) = 0$). In this case, only normalizable modes are present and therefore there is no restriction on the value of the mass aside from the Breitenlohner-Freedman bound. The possible combinations of frequencies that satisfy \eqref{gen res} can be separated into the three distinct cases:
\begin{align}
\oi + \oj + \ok &= \ol \qquad (+++) \\
\oi - \oj - \ok &= \ol \qquad (+--) \\
\oi + \oj - \ok &= \ol \qquad (++-) \, .
\end{align}
Note that the $(+++)$ and $(+--)$ resonances produce restrictions on the allowed values of the indices $\{i, j, k\}$, as well as on values of the mass, since $\oi = 2 i + \Delta^+$. In the first case, the indices are restricted by ${i + j + k = \ell - \Delta^+}$, and so $\Delta^+$ must be an integer and greater than $\ell$ for resonance to occur. Similarly, the $(+--)$ resonance condition becomes ${i - j - k = \ell + \Delta^+}$, which is resonant for any integer value of $\Delta^+$. We will see that these two resonance channels will non-trivially vanish whenever their respective resonance conditions are satisfied. This is in agreement with the results shown by \cite{1407.6273} for a massless scalar in the interior time gauge (as they must be, since the choice of time gauge should not change the existence of resonant channels). Here we include the expressions for the naturally vanishing resonances, choosing to explicitly express the mass dependence. 


\subsection{Naturally Vanishing Resonances: $(+++)$ and $(+--)$}
\label{ssec: zero resonance}

Resonant contributions that come from the condition $\oi + \oj + \ok = \ol$ contribute to the total source term via
\begin{align}
S_\ell = \underbrace{\sum_{i=0}^\infty \sum_{j=0}^\infty \sum_{k=0}^\infty}_{\oi + \oj + \ok = \, \ol} \Omega_{ijk\ell} \, a_i a_j a_k \cos \left( \thi + \thj + \thk \right) + \ldots \, ,
\end{align}
where the ellipsis denotes other resonances from only normalizable modes. $\Omega_{ijk\ell}$ is given by
\begin{align}
\label{omega}
\Omega_{ijk\ell} &= -\frac{1}{12}H_{ijk\ell} \frac{\oj (\oi + \ok +2\oj)}{(\oi + \oj)(\oj + \ok)} - \frac{1}{12} H_{ikj\ell} \frac{\ok (\oi + \oj + 2\ok)}{(\oi + \ok)(\oj + \ok)}- \frac{1}{12} H_{jik\ell} \frac{\oi (\oj + \ok +2\oi)}{(\oi + \oj)(\oi + \ok)} \nonumber \\
& \quad - \frac{m^2}{12} V_{ijk\ell} \left( 1 + \frac{\oj}{\oj + \ok} + \frac{\oi}{\oi + \ok} \right) - \frac{m^2}{12} V_{jki\ell} \left( 1 + \frac{\oj}{\oi + \oj} + \frac{\ok}{\oi + \ok} \right) \nonumber \\
& \quad - \frac{m^2}{12} V_{kij\ell} \left( 1 + \frac{\oi}{\oi + \oj} + \frac{\ok}{\oj + \ok} \right)  + \frac{1}{6} \oj \ok X_{ijk\ell} \left( 1 + \frac{\oj}{\oi + \ok} + \frac{\ok}{\oi + \oj} \right) \nonumber \\
& \quad + \frac{1}{6} \oi \ok X_{jki\ell} \left( 1 + \frac{\oi}{\oj + \ok} + \frac{\ok}{\oi + \oj} \right) + \frac{1}{6} \oi \oj X_{kij\ell} \left( 1 + \frac{\oi}{\oj + \ok} + \frac{\oj}{\oi + \ok} \right) \nonumber \\
& \quad - \frac{1}{12} Z^-_{ijk\ell} \left( \frac{\ok}{\oi + \oj} \right) - \frac{1}{12} Z^-_{ikj\ell} \left( \frac{\oj}{\oi + \ok} \right) - \frac{1}{12} Z^-_{jki\ell}  \left( \frac{\oi}{\oj + \ok} \right) \, .
\end{align}

The second naturally vanishing resonance comes from the condition $\oi - \oj - \ok = \ol$, and contributes to the total source term via
\begin{align}
S_\ell = \sum_{j=0}^\infty \sum_{k=0}^\infty \Gamma_{(j + k + \ell + \Delta^+) jk\ell} \, a_j a_k a_{(j+k+\ell + \Delta^+)} \cos \left( \theta_{(j+k+\ell + \Delta^+)} - \thj - \thk \right) + \ldots \, ,
\end{align}
where
\begin{align}
\label{gamma}
\Gamma_{ijk\ell} &= \frac{1}{4} H_{ijk\ell} \frac{\oj (\ok - \oi + 2\oj)}{(\oi - \oj)(\oj + \ok)} + \frac{1}{4} H_{jki\ell} \frac{\ok (\oj - \oi + 2\ok)}{(\oi - \ok)(\oj + \ok)} + \frac{1}{4} H_{kij\ell} \frac{\oi (\oj + \ok - 2\oi)}{(\oi - \oj)(\oi - \ok)} \nonumber \\
& \quad -\frac{1}{2} \oj \ok X_{ijk\ell} \left( \frac{\ok}{\oi - \oj} + \frac{\oj}{\oi - \ok} - 1\right) + \frac{1}{2} \oi \ok X_{jki\ell} \left( \frac{\ok}{\oi - \oj} + \frac{\oi}{\oj + \ok} - 1 \right) \nonumber \\
& \quad + \frac{1}{2} \oi \oj X_{kij\ell} \left( \frac{\oj}{\oi - \ok} + \frac{\oi}{\oj + \ok} -1 \right) + \frac{m^2}{4} V_{jki\ell} \left( \frac{\oj}{\oi - \oj} + \frac{\ok}{\oi - \ok} -1\right) \nonumber \\
& \quad - \frac{m^2}{4} V_{kij\ell} \left( \frac{\oi}{\oi - \oj} + \frac{\ok}{\oj + \ok} + 1\right) - \frac{m^2}{4} V_{ijk\ell} \left( \frac{\oi}{\oi - \ok} + \frac{\oj}{\oj + \ok} + 1 \right) \nonumber \\
& \quad + \frac{1}{4} Z^-_{kji\ell} \left( \frac{\oi}{\oj + \ok}\right) - \frac{1}{4} Z^+_{ijk\ell} \left( \frac{\ok}{\oi - \oj} \right) - \frac{1}{4} Z^+_{jki\ell} \left( \frac{\oj}{\oi - \ok}\right) \, .
\end{align}

Building on the work done with massless scalars, we are able to show numerically that \eqref{omega} and \eqref{gamma} continue to vanish for massive scalars ($m^2 \geq m^2_{BF}$) in the boundary gauge, in agreement with work done by \cite{1810.04753} in the interior gauge. Thus, the dynamics governing the weakly turbulent transfer of energy are determined only from the remaining resonance channel. Since these resonances vanish, they are not included in the expression for the total source term in~\S\!~\ref{subs: ttf resonances}.


\subsection{Non-vanishing Resonance: $(++-)$}
\label{subs: ttf resonances}

The first non-vanishing contributions arise when $\oi + \oj = \ok + \ol$. This contribution can be split into three coefficients that are evaluated for certain subsets of the allowed values for the indices, namely
\begin{align}
S_\ell &= T_\ell a^3_\ell \cos (\thl + \thl - \thl) + \sum_{i \neq \ell}^\infty R_{i \ell} \, a^2_i a_\ell \cos(\thi + \thl - \thi) \nonumber \\
& \qquad + \sum_{i \neq \ell}^\infty \sum_{j \neq \ell}^\infty S_{i j (i + j - \ell) \ell} \, a_i a_j a_{(i + j - \ell)} \cos(\thi + \thj - \theta_{i + j -\ell} ) \, ,
\end{align}
where the coefficients are given by
\begin{align}
\label{S_ppm}
S_{ijk\ell} &= - \frac{1}{4} H_{kij\ell} \frac{\oi (\oj - \ok + 2\oi)}{(\oi - \ok)(\oi + \oj)} -\frac{1}{4} H_{ijk\ell} \frac{\oj (\oi - \ok + 2\oj)}{(\oj - \ok)(\oi + \oj)} - \frac{1}{4} H_{jki\ell} \frac{\ok ( \oi + \oj - 2\ok)}{(\oi - \ok)(\oj - \ok)} \nonumber \\
& \quad - \frac{1}{2} \oj \ok X_{ijk\ell} \left( \frac{\oj}{\oi - \ok} - \frac{\ok}{\oi + \oj} + 1 \right) - \frac{1}{2} \oi \ok X_{jki\ell} \left( \frac{\oi}{\oj - \ok} - \frac{\ok}{\oi + \oj} + 1 \right) \nonumber \\
& \quad + \frac{1}{2} \oi \oj X_{kij\ell} \left( \frac{\oi}{\oj - \ok} + \frac{\oj}{\oi - \ok} + 1 \right) - \frac{m^2}{4} V_{ijk\ell} \left( \frac{\oi}{\oi - \ok} + \frac{\oj}{\oj - \ok} + 1\right) \nonumber \\
& \quad + \frac{m^2}{4} V_{jki\ell} \left( \frac{\ok}{\oi - \ok} - \frac{\oj}{\oi + \oj} - 1 \right) + \frac{m^2}{4} V_{kij\ell} \left( \frac{\ok}{\oj - \ok} - \frac{\oi}{\oi + \oj} - 1 \right) \nonumber \\
& \quad + \frac{1}{4}  Z^-_{ijk\ell} \left( \frac{\ok}{\oi + \oj}\right)  + \frac{1}{4}  Z^+_{ikj\ell} \left( \frac{\oj}{\oi - \ok}\right) + \frac{1}{4} Z^+_{jki\ell} \left( \frac{\oi}{\oj - \ok} \right) \, ,
\end{align}
\begin{align}
\label{R_ppm}
R_{i\ell} &= \left(\frac{\oi^2}{\ol^2 - \oi^2} \right) \big( Y_{i\ell \ell i} - Y_{i\ell i \ell} + \ol^2 ( X_{i\ell i \ell} - X_{\ell i \ell i}) \big) + \left(\frac{\oi^2}{\ol^2 - \oi^2}\right) \big( H_{\ell i i\ell} + m^2 V_{ii\ell \ell} - 2\oi^2 X_{\ell i i \ell} \big) \nonumber \\
& - \left(\frac{\ol^2}{\ol^2 - \oi^2} \right) \big( H_{i\ell i \ell} + m^2 V_{\ell i i \ell} - 2\oi^2 X_{i\ell i\ell} \big) - \frac{m^2}{4}(V_{i\ell i \ell} + V_{ii\ell \ell} ) + \oi^2 \ol^2 (P_{ii\ell} - 2P_{\ell \ell i}) \nonumber \\
& - \oi\ol X_{i\ell i \ell} - \frac{3m^2}{2} V_{\ell ii \ell} - \frac{1}{2} H_{ii\ell \ell} + \ol^2 B_{ii\ell} - \oi^2 M_{\ell \ell i} - m^2 \oi^2 Q_{\ell \ell i} \, ,
\end{align}
and
\begin{align}
\label{T_ppm}
T_{\ell} &= \frac{1}{2} \ol^2 \left( X_{\ell \ell \ell \ell} + 4 B_{\ell \ell \ell} -2 M_{\ell \ell \ell} - 2m^2 Q_{\ell \ell \ell} \right) -\frac{3}{4} \left( H_{\ell \ell \ell \ell} + 3m^2 V_{\ell \ell \ell \ell} \right) \, .
\end{align}

\begin{figure}
	\centering
		\includegraphics[width=\textwidth]{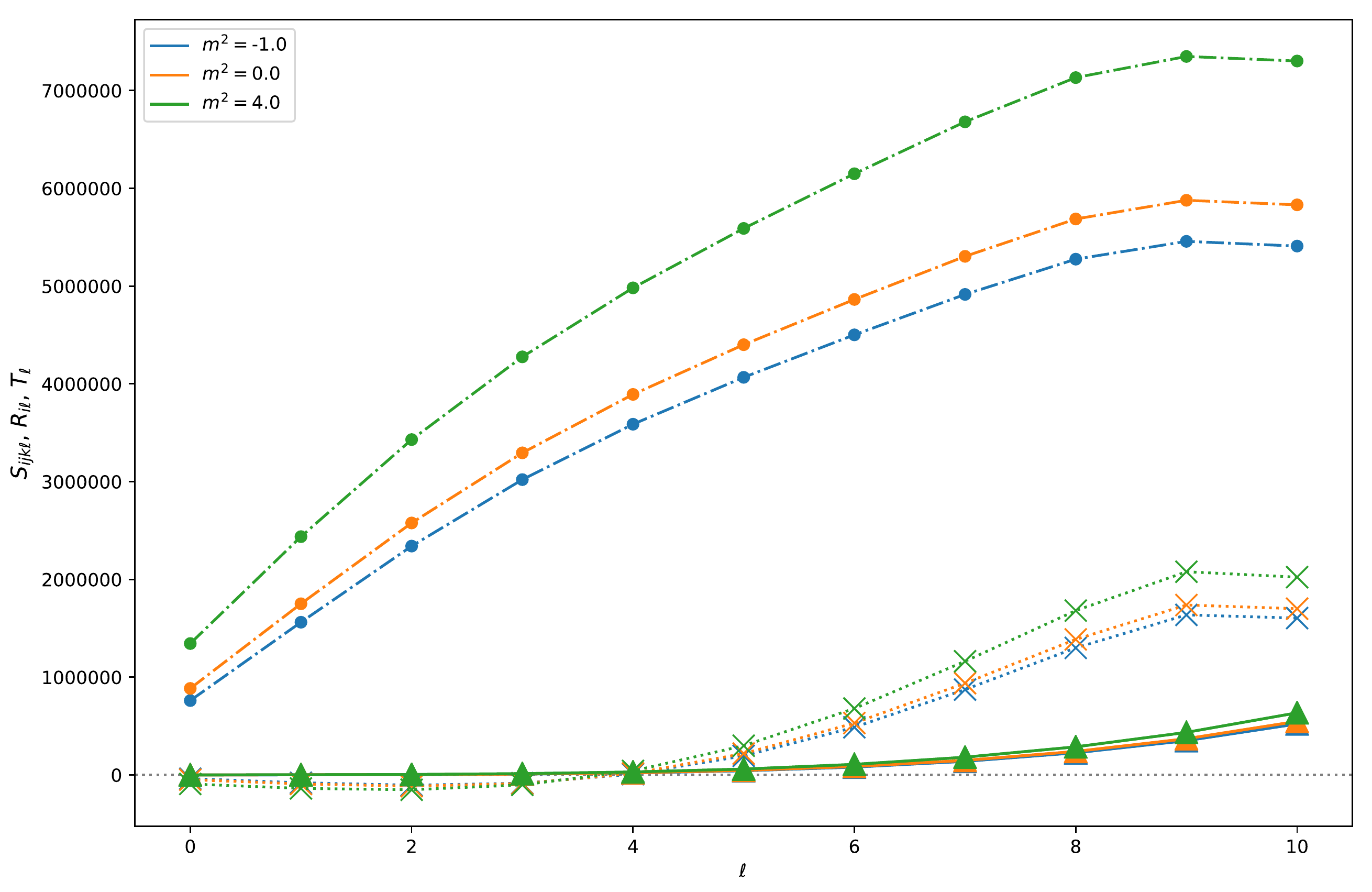}
		\caption{Evaluating \eqref{S_ppm}-\eqref{T_ppm} over different values of $m^2$ for $\ell \leq 10$. $S_{ij(i+j-\ell)\ell}$ is denoted by filled circles connected by dash-dotted lines; $R_{i\ell}$ is denoted by filled triangles connected by solid lines; $T_{\ell}$ is denoted by large Xs connected by dotted lines. Different values of $m^2$ are denoted by the colour of each series.}
		\label{fig: Nmodes}
	\end{figure}

Following the form of \eqref{RN flow 1}~-~\eqref{RN flow 2}, these resonant terms set the evolution of the renormalized integration coefficients to be \cite{1412.3249}
\begin{align}
\frac{2 \ol}{\epsilon^2} \frac{d a_\ell}{d t} &= -  \sum_{i \neq \ell}^\infty \sum_{j \neq \ell}^\infty S_{i j (i + j - \ell) \ell} \, a_i a_j a_{(i + j - \ell)} \sin ( b_\ell + b_{(i+j-\ell)} - b_i - b_j ) \, , \\
\frac{2 \ol a_\ell}{\epsilon^2} \frac{d b_\ell}{d t} &= - T_\ell a_\ell^3 - \sum^\infty_{i \neq \ell} R_{i\ell} \, a_i^2 a_\ell \nonumber \\
& \qquad - \sum_{i \neq \ell}^\infty \sum_{j \neq \ell}^\infty S_{i j (i + j - \ell) \ell} \, a_i a_j a_{(i + j - \ell)} \cos( b_\ell + b_{(i+j-\ell)} - b_i - b_j ) \, .
\end{align}

To examine the effects of non-zero masses on $R$, $S$, and $T$, we evaluate \eqref{S_ppm}-\eqref{T_ppm} for tachyonic, massless, and massive scalars in figure~\ref{fig: Nmodes}. The result is a vertical shift in the coefficient value that is proportional to the choice of mass-squared. By inspection, there is an indication that this shift increases with increasing $\ell$ values; however, a numerical fit of the data would be needed to claim this definitively.


\bibliographystyle{JHEP}
\bibliography{DrivenTTF}


\end{document}